\documentclass[prd,twocolumn,showpacs,nofootinbib,aps]{revtex4-1}
\usepackage{bm}
\usepackage{graphicx}
\usepackage{multirow}
\usepackage{amsmath}
\usepackage{amssymb}
\usepackage{hyperref}
\usepackage{color}
\usepackage{xspace}
\usepackage{verbatim}
\usepackage{mathtools}
\usepackage{booktabs}
\usepackage[dvipsnames]{xcolor}
\usepackage[caption=false]{subfig}

\newcommand{\vmin}{v_\mathrm{min}}

\newcommand{\avg}[1]{\left\langle #1\right\rangle}
\newcommand{\dbd}[2]{\frac{\mathrm{d}#1}{\mathrm{d}#2}}

\begin{document}

\title{Earth-Scattering of super-heavy Dark Matter: \\updated constraints from detectors old and new}
\author{Bradley J. Kavanagh}
\email{b.j.kavanagh@uva.nl}
\affiliation{GRAPPA, Institute of Physics, University of Amsterdam, 1098 XH Amsterdam, The Netherlands}
\affiliation{LPTHE, CNRS, UMR 7589, 4 Place Jussieu, F-75252, Paris, France}

\begin{abstract}
Direct searches for Dark Matter (DM) are continuously improving, probing down to lower and lower DM-nucleon interaction cross sections. For strongly-interacting massive particle (SIMP) Dark Matter, however, the accessible cross section is bounded from above due to the stopping effect of the atmosphere, Earth and detector shielding. We present a careful calculation of the SIMP signal rate, focusing on super-heavy DM ($m_\chi \gtrsim 10^5 \,\,\mathrm{GeV}$) for which the standard nuclear-stopping formalism is applicable, and provide code for implementing this calculation numerically. With recent results from the low-threshold CRESST 2017 surface run, we improve the maximum cross section reach of direct detection searches by a factor of around 5000, for DM masses up to $10^8 \,\,\mathrm{GeV}$. A reanalysis of the longer-exposure, sub-surface CDMS-I results (published in 2002) improves the previous cross section reach by two orders of magnitude, for masses up to $10^{15} \,\,\mathrm{GeV}$. Along with complementary constraints from SIMP capture and annihilation in the Earth and Sun, these improved limits from direct nuclear scattering searches close a number of windows in the SIMP parameter space in the mass range $10^6$ GeV to $10^{13}$ GeV, of particular interest for heavy DM produced gravitationally at the end of inflation.
 
\end{abstract}

\maketitle

\section{Introduction}
\label{sec:introduction}

Direct detection experiments aim to detect Dark Matter (DM) by measuring the energy of nuclear recoils, induced by the interaction of DM in the Galactic halo with nuclei in the detector \cite{Goodman:1984dc, Drukier:1986tm}. As standard direct detection experiments probe smaller and smaller cross sections for weak-scale DM masses  \cite{Aprile:2017iyp,Cui:2017nnn}, low-threshold experiments \cite{Petricca:2017zdp,Angloher:2017sxg} and DM-electron scattering searches \cite{Essig:2017kqs} are beginning to search for DM with GeV-scale masses and below. With no confirmed detection thus far, we are driven to search as much of the DM parameter space as possible. Indeed, there are well-motivated particle DM candidates ranging from the very light \cite{Knapen:2017xzo,An:2014twa} to the very heavy \cite{Kusenko:1997si,Chung:1998ua,Kolb:2007vd,Kolb:2017jvz}, interacting with the Standard Model very weakly \cite{Feng:2003xh,Steffen:2006hw,Benakli:2017whb} or very strongly \cite{Schissel:2006kx,Hochberg:2014dra,Bruggisser:2016ixa}. In this paper, we revisit direct detection constraints on such strongly-interacting DM, with a focus on super-heavy candidates.

There has been much recent interest in signals and constraints for DM which interacts strongly with nuclei and whose distribution may therefore be affected by scattering in the Earth \cite{Collar:1992qc,Collar:1993ss,Hasenbalg:1997hs}. Much of this has been focused on light (sub-GeV) DM where lower limits from conventional direct detection experiments weaken, meaning that moderate DM-nucleon scattering cross sections have not yet been excluded \cite{Foot:2011fh,Kouvaris:2014lpa,Kouvaris:2015laa,Foot:2014osa,Clarke:2015gqw,Bernabei:2015nia,Emken:2017erx,Davis:2017noy}. However, care must be taken in the treatment of DM scattering in the Earth. Light DM may be deflected substantially when it scatters with nuclei, drastically changing the incoming direction and path length \cite{Kavanagh:2016pyr,Mahdawi:2017cxz,Mahdawi:2017utm}. However, the recently developed \textsc{DaMaSCUS} Monte Carlo code \cite{Emken2017a} accounts for these effects and can be used to reliably predict signals of strongly-interacting light DM over a range of cross sections \cite{Emken:2017qmp}.

Heavier DM, on the other hand, is already strongly constrained by direct detection experiments. But constraints typically cut off at large values of the DM-nucleon cross section, where scattering in the Earth slows the DM particles before they reach the detector \cite{Starkman:1990nj,Albuquerque:2003ei}, rendering them undetectable. At even larger cross sections, airborne experiments provide constraints, as the less dense atmosphere is not as effective at stopping DM particles \cite{Davis:2017noy}. A number of small windows remain open between these constraints for DM masses larger than $\sim10^5$ GeV, in the mass range favoured by gravitational production at the end of inflation \cite{Hui:1998dc,Allahverdi:2002nb,Kannike:2016jfs}. Though these regions are disfavoured by complementary constraints from considering DM capture and annihilation in the Earth and Sun \cite{Mack:2007xj,Albuquerque:2010bt,Mack:2012ju}, it is important to cross-check using direct searches, which rely on an entirely different set of assumptions. In this work, we demonstrate that the maximum cross section excluded by existing direct detection experiments can be increased by more than three orders of magnitude, excluding these remaining windows using nuclear scattering experiments alone. We also provide a numerical code, \textsc{verne} \cite{verne}, which can be used to calculate the Earth-stopping effect for heavy DM and derive lower limits on the DM-nucleon cross section.

We focus here on two experiments: the CRESST 2017 surface run \cite{Angloher:2017sxg}, operated at the Max-Planck-Institute for Physics (MPI) Munich in 2017, and CDMS-I \cite{Abusaidi:2000wg,Abrams:2002nb}, operated at the Stanford Underground Facility (SUF) in 1998-1999. The CRESST 2017 surface run provides excellent sensitivity to strongly interacting massive particles (SIMPs) because of its low threshold of 20 eV, as well as the fact that it was operated in a surface building, meaning that there is limited shielding by the Earth. We extend the SIMP limits presented in Ref.~\cite{Davis:2017noy} to large masses. However, the short  ($\sim$ few hours) exposure of the CRESST 2017 surface run means that it is not sensitive to DM masses larger than $\sim10^8$ GeV, when the number of DM particles crossing the detector over the course of the exposure becomes too small to generate an appreciable signal. Instead, the CDMS-I experiment had a longer ($\sim$ 1 year) exposure and therefore probes masses up to  $\sim10^{15}$ GeV. Limits on SIMPs from CDMS-I were derived in Ref.~\cite{Albuquerque:2003ei} and we present a refinement of those limits, taking into account form factor suppression of the SIMP scattering underground, as well as modelling the full (anisotropic) velocity distribution of incoming SIMPs.

In Sec.~\ref{sec:DirectDetection}, we present the general framework of direct detection, then in Sec.~\ref{sec:NuclearStopping}, we describe the formalism for calculating the final DM velocity at the detector, after propagation through the atmosphere, Earth and detector shielding. We focus on heavy DM ($m_\chi \gtrsim 10^5 \,\,\mathrm{GeV}$) with large scattering cross sections, which allows us to make two simplifications. First, DM with a cross section $\sim 10^{-28}\,\,\mathrm{cm}^2$ would typically scatter $N_\mathrm{scat}\sim\mathcal{O}(5000)$ times before reaching the CDMS-I detector at SUF. Thus, we can approximate the energy losses as continuous over the path of the particles. In addition, for each scatter with a nucleus of mass $m_N$, the DM is deflected by an angle $\delta \theta \sim m_N/m_\chi \lesssim 10^{-3}$ \cite{Kavanagh:2016pyr}. The total deflection of the DM particle is then expected to be $\Delta \theta \sim \delta \theta/\sqrt{N_\mathrm{scat}} \lesssim \mathcal{O}(10^{-5})$. We can thus approximate the trajectories of DM particles as straight lines, allowing us to use the standard nuclear stopping formalism and avoiding the need for complicated Monte Carlo calculations.

With this formalism in hand, we derive the final velocity distribution at the detectors in Sec.~\ref{sec:VelDist}, followed by the expected signal rate and resulting constraints in Sec.~\ref{sec:constraints}. The final constraints are shown in Fig.~\ref{fig:constraints}, for DM masses in the range $1$ GeV to $10^{15}$ GeV, with the caveat that at low masses these results need to be supplemented with dedicated Monte Carlo simulations. All code for performing the calculations in this paper, along with all results and plots, is made freely available online, along with the numerical code \textsc{verne}, at \href{https://github.com/bradkav/verne}{https://github.com/bradkav/verne} \cite{verne}.

\section{Direct detection of SIMPs}
\label{sec:DirectDetection}

In order to determine the maximum cross section probed by a particular experiment, we must compare the number of observed events with the number of expected SIMP signal events. For DM scattering with a nucleus $N$, the rate of nuclear recoils of energy $E_R$ per unit detector mass is given by \cite{Jungman:1995df},
\begin{align}
\begin{split}
\label{eq:dRdE}
\frac{\mathrm{d}R}{\mathrm{d}E_R} = \frac{\rho_\chi}{m_\chi}   \int_{\vmin}^\infty v f(\mathbf{v}) \dbd{\sigma_{\chi N}}{E_R} \, \mathrm{d}^3\mathbf{v}\,.
\end{split}
\end{align}
Here, $\rho_\chi$ is the local DM density, which we fix to the benchmark value of $0.3\,\,\mathrm{GeV} \,\,\mathrm{cm}^{-3}$ \cite{Green:2011bv}, noting that there is some uncertainty on the true value \cite{Read:2014qva}. For the differential DM-nucleus cross section $\mathrm{d}\sigma_{\chi N}/\mathrm{d}E_R$ we assume standard spin-independent (SI) scattering. We integrate over the DM velocity distribution $f(\mathbf{v})$, including the contribution of all particles with sufficient DM speed to excite a recoil of energy $E_R$. That is, we include all particles with $v > v_\mathrm{min} = \sqrt{m_N E_R/(2 \mu_{\chi N}^2)}$, where $\mu_{\chi N}$ is the DM-nucleus reduced mass. 

For weakly interacting massive particles (WIMPs), the velocity distribution is typically assumed to have the form:\footnote{We distinguish between the velocity $\mathbf{v} = (v_x, v_y, v_z)$ of the DM particles and their speed $v = |\mathbf{v}|$. The speed distribution $f(v)$ is related to the velocity distribution $f(\mathbf{v})$ by $f(v) = \int v^2 f(\mathbf{v}) \,\mathrm{d}^2\hat{\mathbf{v}}$.}
\begin{align}
\begin{split}
\label{eq:veldist}
f(\mathbf{v}) \propto \exp\left(-\frac{\left|\mathbf{v} - \langle \mathbf{v}_\chi \rangle\right|^2}{2 \sigma_v^2}\right) \Theta(v_\mathrm{esc} - \left|\mathbf{v} - \langle\mathbf{v}_\chi\rangle\right|)\,.
\end{split}
\end{align}
This velocity distribution defines the so-called Standard Halo Model (SHM), for which we assume an escape speed in the Galactic frame $v_\mathrm{esc} = 533 \,\,\mathrm{km/s}$ \cite{Piffl:2013mla} and dispersion $\sigma_v = v_\mathrm{circ}/\sqrt{2} \approx 156 \,\, \mathrm{km/s}$ \cite{Green:2011bv}. The mean DM velocity is given by $\langle \mathbf{v}_\chi \rangle = - \mathbf{v}_\mathrm{lab}(t)$, the velocity of the lab with respect to the Galactic rest frame. The lab velocity is time-varying, arising from the motion of the Sun around the Galactic centre, the orbit of the Earth around the Sun as well as the daily revolution of the Earth \cite{Bozorgnia:2011tk,Mayet:2016zxu}. 

Equation~\eqref{eq:veldist} corresponds to the initial velocity distribution of DM particles arriving at Earth and will be modified if DM undergoes substantial interactions in the atmosphere or in the Earth itself. This modification depends on the path travelled by the DM particles or, equivalently, the position of the detector on Earth with respect to the DM flux. It is useful to define an angle $\gamma$, defined as the angle between the position vector of the detector $\mathbf{r}_\mathrm{det}$ (as measured from the centre of the Earth) and the direction of the mean DM flux:
\begin{equation}
\label{eq:gamma}
\gamma = \cos^{-1}\left(\langle \hat{\mathbf{v}}_\chi\rangle \cdot \hat{\mathbf{r}}_\mathrm{det}\right)\,.
\end{equation}
With this definition, the mean DM flux appears to come from directly overhead for $\gamma = 180^\circ$ while the flux appears to come from below (through most of the Earth) for $\gamma = 0^\circ$. Explicit expressions for the velocity distribution in terms of the angle $\gamma$ are given in Appendix~\ref{app:coordinates}.

In Fig.~\ref{fig:gamma}, we plot the value of $\gamma$ for the Stanford Underground Facility (SUF, latitude $37.4^\circ \mathrm{N}$, longitude $122.2^\circ \mathrm{W}$) for a 1-year period covering the CDMS-I exposure presented in Ref.~\cite{Abusaidi:2000wg,Abrams:2002nb}. The slow variation due to the Earth's orbit and the fast oscillations from the Earth's rotation are both visible. On average, the flux of DM appears to come from above, at an angle $\sim 53^\circ$ off the vertical. This suggests that the typical path length for DM particles arriving at SUF will be relatively short, as the particles do not need to cross the entire Earth to reach the detector. The picture is similar for the Max Planck Institute in Munich (MPI, latitude $48.1^\circ \mathrm{N}$, longitude $11.57^\circ \mathrm{E}$), owing to the similar Northern latitude.

\begin{figure}[t]
\centering
\includegraphics[width=0.49\textwidth,]{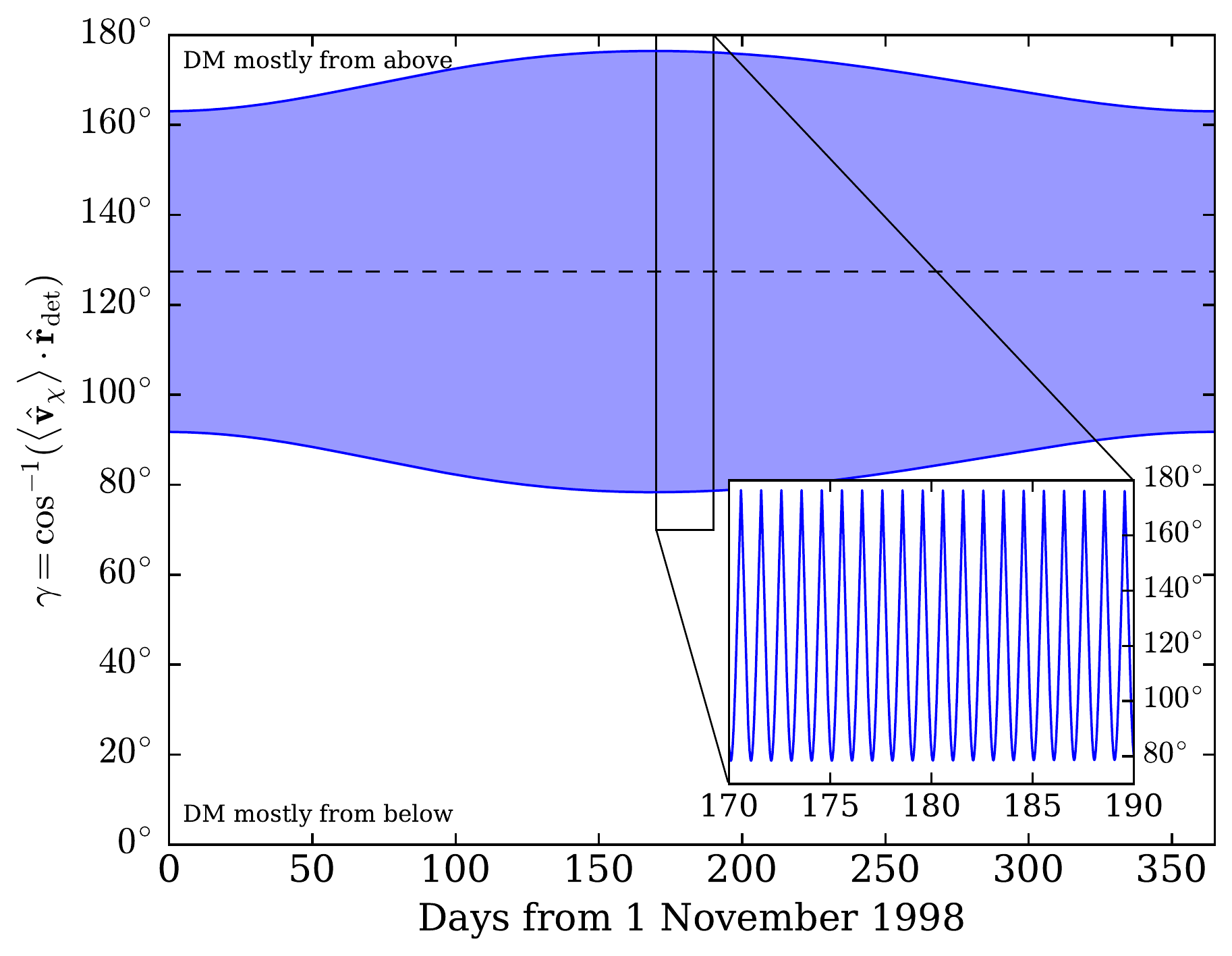}
\caption{\textbf{Average direction of incoming Dark Matter flux at the Stanford Underground Facility (SUF).}  The average direction is described by the angle $\gamma$ defined in Eq.~\eqref{eq:gamma} and the surrounding text. The shaded blue band shows the maximum and minimum values over the course of the year while the zoomed inset shows the daily variation due to the Earth's rotation.}
\label{fig:gamma}
\end{figure}

The variation of $\gamma$ during the exposure means that DM particles will travel different path lengths through the atmosphere and Earth at different times. This in turn leads to a variation in the stopping effect and a time variation in the velocity distribution $f(\mathbf{v}, \gamma(t))$ of particles reaching the detector. The (time-varying) rate of recoils in the detector is then:
\begin{align}
\begin{split}
\dbd{N_e}{t} = M \int_{E_\mathrm{min}}^{E_\mathrm{max}} \dbd{R(t)}{E_R} \epsilon(E_R) \,\mathrm{d}E_R\,,
\end{split}
\end{align}
where $M$ is the fiducial mass of the experiment, $\epsilon(E_R)$ is the nuclear recoil detection efficiency\footnote{Here, we assume that any energy resolution effects have been folded into the efficiency function.} and the analysis window of the experiment is $E_R \in [E_\mathrm{min}, E_\mathrm{max}]$. In the case of the CRESST 2017 surface run, we calculate the expected number of signal events by integrating over the full 5.31 hour exposure on the night of 16th February 2017 (see the ancillary material to Ref.~\cite{Angloher:2017sxg} for more details). For CDMS-I, we do not know the exact times of the exposure, so we integrate the event rate over the course of a typical day, then multiply by the total number of exposure days:
\begin{equation}
\label{eq:Nexp}
N_e = N_\mathrm{days} \int_{\mathrm{1\,\,day}} \left.\dbd{N_e}{t}\right|_{ \gamma(t)} \,\mathrm{d}t\,.
\end{equation}
We select this typical day as 21st April 1999 (somewhere mid-way through the full exposure). 

While $\gamma = 180^\circ$ leads to the shortest path length for incident DM particles and therefore the largest rate, the detectors will generally not spend a long time oriented such that $\gamma = 180^\circ$. It is therefore important to calculate the speed distributions for a range of values of $\gamma$ and take all of these into account. In the next two sections, we explain how to calculate the DM velocity at the detector for a given trajectory, followed by the procedure for evaluating the speed distribution at the detector, accounting for all possible trajectories.


\section{Nuclear stopping of Dark Matter}
\label{sec:NuclearStopping}

In order to determine the final velocity of DM particles at the detector, we use the `nuclear stopping' approach, first presented in Ref.~\cite{Starkman:1990nj}. Our treatment follows closely a number of references, including Ref.~\cite{Davis:2017noy}. However, there are a number of refinements to the standard picture which we will make particular note of. For now, we keep the DM mass $m_\chi$ general, before focusing on the case of super-heavy DM ($m_\chi \gg m_A$).

As DM particles traverse a medium, consisting of different nuclear species which we label with the index $i$, they may interact and lose energy. As we argued in the introduction, we can assume that this scattering is continuous, meaning that the rate of change of the average DM energy is given by:
\begin{equation}
\label{eq:energyloss}
\dbd{\langle E_\chi \rangle}{t} = -\sum_{i} n_i(\mathbf{r}) \, \langle E_R \rangle_i \,\sigma_i(v) \, v\,,
\end{equation} 
where $n_i(\mathbf{r})$ is the number density of nuclei of species $i$ at position $\mathbf{r}$, $\sigma_i(v)$ is the total DM-nucleus scattering cross section and $v$ is the DM speed. The average recoil energy (or equivalently the average change in DM energy) for scattering with nucleus $i$ is given by:
\begin{align}
\langle E_R \rangle_i&= \frac{1}{\sigma_i(v)}\int_{0}^{E^{\mathrm{max}}_i} E_R \dbd{\sigma_i}{E_R} \, \mathrm{d}E_R\,.
\end{align}
Here, $E_i^\mathrm{max} = 2 \mu_{\chi i}^2 v^2/m_i$ is the maximum recoil energy which can be transmitted to nucleus $i$ through elastic scattering.

For standard spin-independent (SI) scattering, the differential cross section is given by \cite{Lewin:1995rx,Cerdeno:2010jj}:
\begin{align}
\dbd{\sigma_i}{E_R} &= \frac{m_i \sigma_p^{\mathrm{SI}}}{2 \mu_{\chi p}^2 v^2} A_i^2 F_i^2(E_R)\,.
\end{align}
The strength of the interaction is parametrised by the DM-proton cross section at zero momentum transfer $\sigma_p^{SI}$ and the total cross section is coherently enhanced by the number of nucleons in the target nucleus, $A_i^2$. The form factor $F_i^2(E_R)$ (which we take to have the Helm form \cite{Helm:1956zz,Duda:2006uk}) arises due to the finite size of the nucleus and suppesses the cross section at large recoil energies. For the SI interaction, the mean energy loss is then given by:
\begin{align}
\begin{split}
\sigma_i (v) \avg{E_R}_i &= \frac{\mu_{\chi i}^4 v^2}{\mu_{\chi p}^2 m_i} \sigma_p^{\mathrm{SI}}A_i^2 \int_0^1 (2x)\,F_i^2(x E_i^\mathrm{max})\, \mathrm{d}x\\
&=\frac{\mu_{\chi i}^4 v^2}{\mu_{\chi p}^2 m_i} \sigma_p^{\mathrm{SI}}A_i^2 C_i(m_\chi, v)\,.
\end{split}
\end{align}

The coherence factor $C_i(m_\chi, v)$ reflects the suppression of the mean recoil energy (relative to that expected for a point-like nucleus) owing to the nuclear form factor. For very light DM (and for point-like nuclei) $C_i(m_\chi, v) \rightarrow 1$ meaning that this factor is typically neglected in studies of the Earth-stopping of sub-GeV mass DM \cite{Davis:2017noy}. For heavy DM and heavy target nuclei, however, the effects of the form factor can be large. For a DM particle of mass $10^5 \,\mathrm{GeV}$ scattering off lead, for example, $C_\mathrm{Pb}(m_\chi, v)$ drops below 1\% for DM speeds larger than around 200 km/s. This reduces the shielding efficiency of lead (and the stopping efficiency of Earth elements) for fast, heavy DM. It is therefore imperative that we include it here.

We now examine in more detail the argument given in the introduction that the scattering can be described as continuous and that ultra-heavy DM particles should travel in a straight line.  The typical number of scattering events which a DM particle experiences is:
\begin{align}
\label{eq:Nscat}
 N_\mathrm{scat} = \sum_i n_i \sigma_i L \approx 500 \,\left(\frac{\sigma_p^\mathrm{SI}}{10^{-28}\,\mathrm{cm}^2}\right)\left(\frac{D}{\mathrm{m}}\right)\,,
\end{align}
where $D$ is the distance travelled. This corresponds to a mean free path of $\lambda \approx 2 \,\mathrm{mm}$. We have taken here typical number densities in the Earth's crust and assumed that the DM is much heavier than the most abundant elements in the crust. For a detector $10 \,\mathrm{m}$ underground, DM particles with a cross section of $\sigma_p^{\mathrm{SI}} = 10^{-28}\,\mathrm{cm}^2$ will scatter around 5000 times with a roughly gaussian error of $\sqrt{N_\mathrm{scat}} \sim 70$. Treating the stopping as continuous (rather than as a series of discrete scatters) should therefore introduce an error of $\mathcal{O}(1\%)$. As we will see, we are able to exclude DM particles with even larger cross sections that this, in which case the approximation only improves.

For very light DM particles, the angle $\alpha$ (measured in the lab frame)  by which their path is deflected after a scatter is uniform. As we increase the DM mass, however, $\alpha$ becomes increasingly peaked in the forward direction (see Sec.~3.3.3 of Ref.~\cite{Kavanagh:2016pyr} for a more detailed discussion). In the heavy DM limit, we find that $\alpha \in [0, m_i/m_\chi]$. Over many scatters, the typical total deflection of the particle is $\sqrt{\langle\Delta x^2\rangle} \sim \sqrt{N_\mathrm{scat}} (m_i/m_\chi)  \lambda$, corresponding to an angular deflection of $\Delta \alpha \sim \sqrt{\langle\Delta x^2\rangle}/D \sim (m_i/m_\chi)/\sqrt{N_\mathrm{scat}}$. For the case of a $10^{5}\,\mathrm{GeV}$ particle, scattering 5000 times, this corresponds to a typical angular deflection of $\mathcal{O}(10^{-5})$ radians (or a $0.1\,\mathrm{mm}$ deflection over $10 \,\mathrm{m}$ of travel). With increasing DM mass, the typical angular deflection becomes smaller and smaller. In light of this, we assume from now on that the DM particles travel in straight line trajectories.

Writing the DM energy as $E_\chi = m_\chi v^2/2$, we can rewrite Eq.~\eqref{eq:energyloss} in terms of the DM speed $v$ and the straight-line distance travelled by the DM particle $D$:
\begin{align}
\begin{split}
\label{eq:stopping}
\dbd{v}{D} &= -\frac{v}{m_\chi \mu_{\chi p}^2} \sigma_p^{\mathrm{SI}}\sum_i^{\mathrm{species}} n_i (\mathbf{r}) \frac{\mu_{\chi i}^4}{m_i} A_i^2 C_i(m_\chi, v)\\
&\approx -m_p v \left(\frac{\sigma_p^{\mathrm{SI}}}{m_\chi}\right) \sum_i^{\mathrm{species}} n_i (\mathbf{r}) A_i^5 C_i(m_\chi \rightarrow \infty,v)\,,
\end{split}
\end{align}
where on the second line we have taken the limit $m_\chi \gg m_i$, for all target nuclei of interest. For the numerical results of the paper, we use the full expression on the top line of Eq.~\eqref{eq:stopping}, but it is useful to note that for very heavy DM, the stopping effect scales as $\sigma_p^{\mathrm{SI}}/m_\chi$. This means that once we have calculated the final velocity distribution at the detector for a given (large) DM mass and cross section, the results for another DM mass can be obtained trivially by rescaling the cross section appropriately.


Because of the non-trivial dependence of the coherence factor $C_i(m_\chi, v)$ on the DM speed, it is not possible to solve Eq.~\eqref{eq:stopping} analytically. Instead, it must be solved numerically, starting from some initial speed $v_i$ and integrating forward along the trajectory of the DM particle to obtain the final speed at the detector $v_f$. Each DM particle is propagated through three regions, in order:
\begin{itemize}
\item Atmosphere - particles are propagated along a straight line, towards the detector, from the top of the atmosphere to the surface of the Earth. We include stopping by Oxygen and Nitrogen nuclei and take the atmospheric density profiles from the International Standard Atmosphere \cite{ISA}, which extends up to a height of 80 km.
\item Earth - particles then propagate in a straight line from the surface of the Earth to the detector. We include 8 different Earth elements - O, Si, Mg, Fe, Ca, Na, S, Al - using density profiles tabulated in Ref.~\cite{Lundberg:2004dn} (based on data from Refs.~\cite{Geochemistry,Britannica}).
\item Shielding - finally, the particles propagate through any shielding which surrounds the detector. In the case of the atmosphere and the Earth, the path length depends on the angle of incidence, while for the shielding we assume a constant path length, regardless of incoming direction. 
\end{itemize}
Explicit expressions for the path length through the atmosphere and Earth are given in Appendix~\ref{app:coordinates}.

For the CDMS-I detector as SUF, we assume that the detector is located 10.6m underground and is surrounded by 16cm of lead shielding\footnote{Polyethylene and copper shields are also employed, but we neglect these owing to lead's much larger scattering cross section with DM particles.}. For the CRESST 2017 surface run at MPI, the only substantial shielding comes from the walls of the surface building itself, consisting of 30cm of concrete. We model this by assuming that the detector is positioned at a depth of 30 cm underground. In addition, we include shielding from 1mm of copper surrounding the detector.

\begin{figure*}[t]
\centering
\includegraphics[width=0.49\textwidth,]{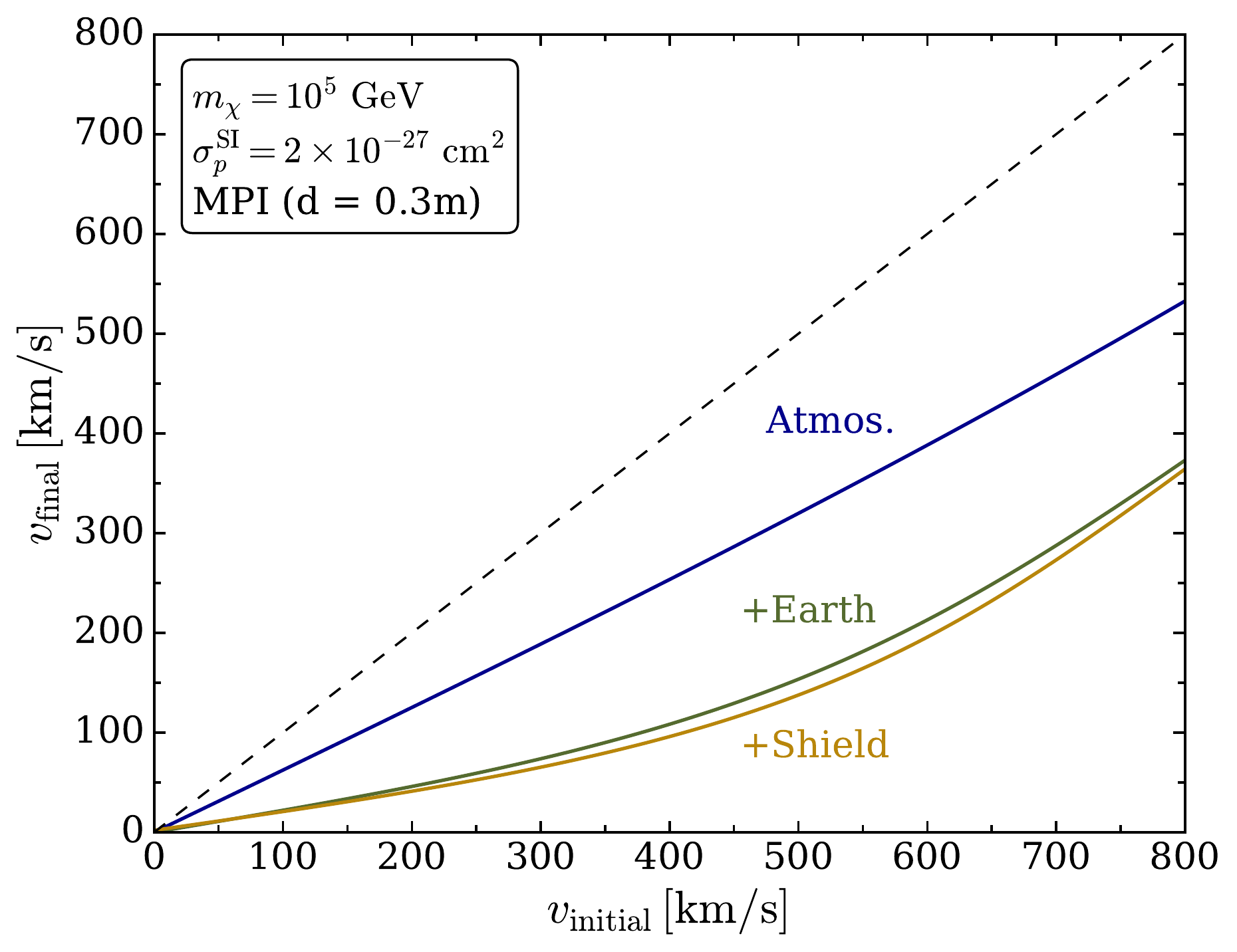}
\includegraphics[width=0.49\textwidth,]{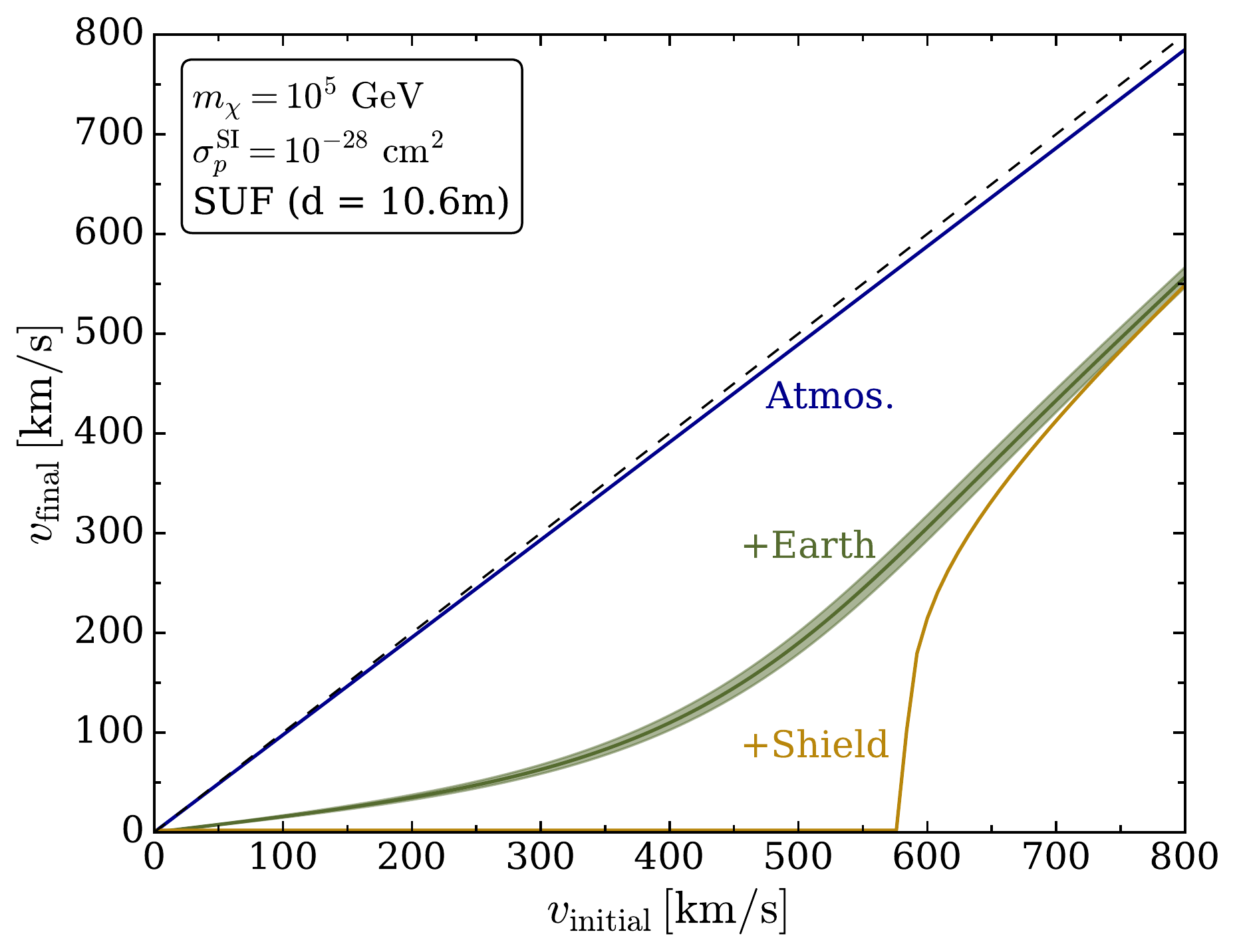}
\caption{\textbf{Example of mapping the initial speed of DM particles to the final speed at the detector.} The dashed line denotes the DM speed at the top of the atmosphere and solid lines denote the final DM speed after propagation through the atmosphere (blue), then after propagation through the Earth (green), then after propagation through the shielding (orange). In this example, we assume DM particles travelling vertically downwards. \textbf{Left:} detector at MPI at a depth of $\sim 30$ cm (CRESST 2017 surface run). \textbf{Right:} detector at SUF  at a depth of 10.6m (CDMS-I). Note the different cross sections in the two cases. The shaded green band in the right panel corresponds to a 4\% uncertainty on the depth of the detector (roughly equivalent to a $3\sigma$ fluctuation on the number of underground scatters).}
\label{fig:VelocityTransfer}
\end{figure*}


In Fig.~\ref{fig:VelocityTransfer}, we show an example of the final DM speed $v_\mathrm{final}$ as a function of the initial DM speed $v_\mathrm{initial}$ for a particular trajectory, namely particles travelling vertically downwards from the top of the atmosphere to the detector. In the left panel, we show results for a detector at SUF and in the right panel for a detector in a surface building at MPI. In both cases, we assume a DM mass of $10^5\,\, \mathrm{GeV}$, but note that in the left panel the DM-nucleon cross section is roughly 20 times larger than in the right.

For a detector at MPI (left panel), we note that the atmospheric crossing (blue line) has the largest slowing effect, as described in Ref.~\cite{Davis:2017noy}. Crossing the 30 cm of concrete which makes up the walls of the surface building (green line) also gives a reduction in speed. Neglecting the coherence factor $C_i(m_\chi, v)$ in Eq.~\eqref{eq:stopping}, we would expect $v_\mathrm{final} \propto v_\mathrm{initial}$, as is usually assumed when considering nuclear stopping. Above around 400 km/s, however, the final speed after crossing the Earth is higher than expected from this simple proportionality. This is due to the coherence factor $C_i(m_\chi, v)$. More energetic particles typically impart much less than the maximum possible recoil energy, especially when scattering with heavy nuclei. Finally, the particles must cross 1mm of copper (orange line), which gives only a small reduction in the speed.

 
 For the case of a detector at SUF (right panel), we consider a smaller DM-nucleon cross section, meaning that the atmosphere (blue line) has only a small effect on the slowing of DM.  Instead, the DM particles must cross 10.6m of earth to reach the detector, leading to substantial slowing at low speeds (green line). Again, high speed particles are slowed less and this is particularly noticeable after the particles have also crossed the lead shielding (orange line), where only the fastest-moving DM particles survive. Neglecting the coherence factor, we would expect that these fast-moving DM particles would be effectively stopped and would not be detectable in an underground experiment.
 
In this section, we have assumed that the energy losses of the particles are continuous and are well-described by the \textit{average} number of scatters. Indeed, some particles will undergo more or fewer scatters than this average, introducing some variation into the final velocity of the particles. For a particle scattering around 5000 times (see Eq.~\eqref{eq:Nscat}), the typical error on the number of scatters is $\mathcal{O}(1\%)$. We can accommodate this into our formalism as an uncertainty on the distance travelled by the particle $D \sim N_\mathrm{scat} \lambda$. In the right panel of Fig.~\ref{fig:VelocityTransfer}, we include a shaded green band, showing the effect of varying the detector depth by $\pm 4\%$ (or roughly a $3\sigma$ variance in the number of underground scatters). The resulting error on the DM velocity is around 5\%, further supporting the assumption that the energy losses can be treated as continuous in this many-scatter regime.



\section{Velocity distribution at the detector}
\label{sec:VelDist}
We have so far discussed the propagation of a given DM particle to the detector. Now we examine the impact on the total population of DM particles to obtain the final speed distribution at the detector\footnote{We could also explore the full 3-dimensional \textit{velocity} distribution at the detector, which would be relevant for directional detectors. However, we leave this for future work.}. We assume that the initial velocity distribution $f(\mathbf{v}_i)$ (at the upper atmosphere, before scattering) has the Maxwell-Boltzmann form given in Eq.~\eqref{eq:veldist}. The final distribution of velocities at the detector $\tilde{f}(\mathbf{v}_f)$ is obtained by a change of variables:
\begin{align}
\begin{split}
\tilde{f}(\mathbf{v}_f) \, \mathrm{d}^3 \mathbf{v}_f &= f(\mathbf{v}_i) \, \mathrm{d}^3 \mathbf{v}_i\\
\Rightarrow  \tilde{f}(\mathbf{v}_f) v_f^2 \,\mathrm{d}v_f \,\mathrm{d}\hat{\mathbf{v}}_f^2 &= f(\mathbf{v}_i) v_i^2\,\mathrm{d}v_i \,\mathrm{d}\hat{\mathbf{v}}_f^2\\
\Rightarrow \tilde{f}(\mathbf{v}_f)  &= f(\mathbf{v}_i) \left(\frac{v_i^2}{v_f^2}\right) \,\dbd{v_i}{v_f}\,,
\end{split}
\end{align}
where in passing to the last line we have used that $\hat{\mathbf{v}}_f = \hat{\mathbf{v}}_i \equiv \hat{\mathbf{v}}$, because the DM particles are assumed to travel in straight lines.

The speed distribution at the detector is then:
\begin{align}
\begin{split}
\label{eq:finaldist}
\tilde{f}(v_f) &\equiv v_f^2 \oint \tilde{f}(\mathbf{v}_f) \, \mathrm{d}^2\hat{\mathbf{v}}\\
&= \oint f(\mathbf{v}_i) v_i^2 \dbd{v_i}{v_f} \, \mathrm{d}^2\hat{\mathbf{v}}\,.
\end{split}
\end{align}
Here, we understand $v_i$ to be the initial speed required to obtain a final speed of $v_f$ for particles travelling along the trajectory specified by the direction $\hat{\mathbf{v}}$. We note also that $v_i$ and the derivative $\mathrm{d}v_i/\mathrm{d}v_f$ depend on the incoming DM direction (and so must be included inside the integral over incoming angles). Evaluating the final speed distribution $\tilde{f}(v_f)$ at the detector then requires us to integrate Eq.~\eqref{eq:stopping} \textit{backwards} from the detector to the top of the atmosphere. 

An alternative approach would be to evaluate $\tilde{f}(v_f)$ by a full Monte Carlo simulation, following the trajectories of a large number of particles between individual scattering events. However, given the large number of scatters expected, $\mathcal{O}(1000)$ or more, such simulations could be prohibitively slow. A simplified Monte Carlo simulation, as performed in Ref.~\cite{Davis:2017noy}, is also possible, in which the final distribution of velocities is found from generating a sample of initial velocities and solving Eq.~\eqref{eq:stopping} for each. From Eq.~\eqref{eq:finaldist}, we see that $\tilde{f}(v_f)$ is typically dominated by trajectories where $\mathrm{d}v_i/\mathrm{d}v_f$ is large. The `back-propagation' method we use automatically takes care of the proper weighting by the derivative $\mathrm{d}v_i/\mathrm{d}v_f$ (which is calculated numerically), thus minimising the number of trajectories which must be evaluated.

For given values of $m_\chi$, $\sigma_p^\mathrm{SI}$ and $\gamma$, we evaluate $\tilde{f}(v_f)$ over a range of values for $v_f$, from 1 km/s up to $v_\mathrm{max}$, the maximum speed which a DM particle can have when it reaches the detector (maximised over all incoming directions). As a cross-check, we can verify that this procedure maintains the normalisation of the distribution function. We integrate $\tilde{f}(v_f)$ in the range $v_f \in [1 \,\,\mathrm{km/s}, \,v_\mathrm{max}]$ and add to this the fraction of the initial particles which have a speed lower than $1 \,\,\mathrm{km/s}$ when they reach the detector\footnote{The initial velocity which gives a final speed $v_f$ \textit{less than} 1 km/s is obtained by back-propagating particles of $v_f = 1\,\,\mathrm{km/s}$ for a range of incoming directions.}. We find that the distribution is always normalised to unity, typically to better than $1$ part in $10^3$.


\begin{figure}[t]
\centering
\includegraphics[width=0.49\textwidth,]{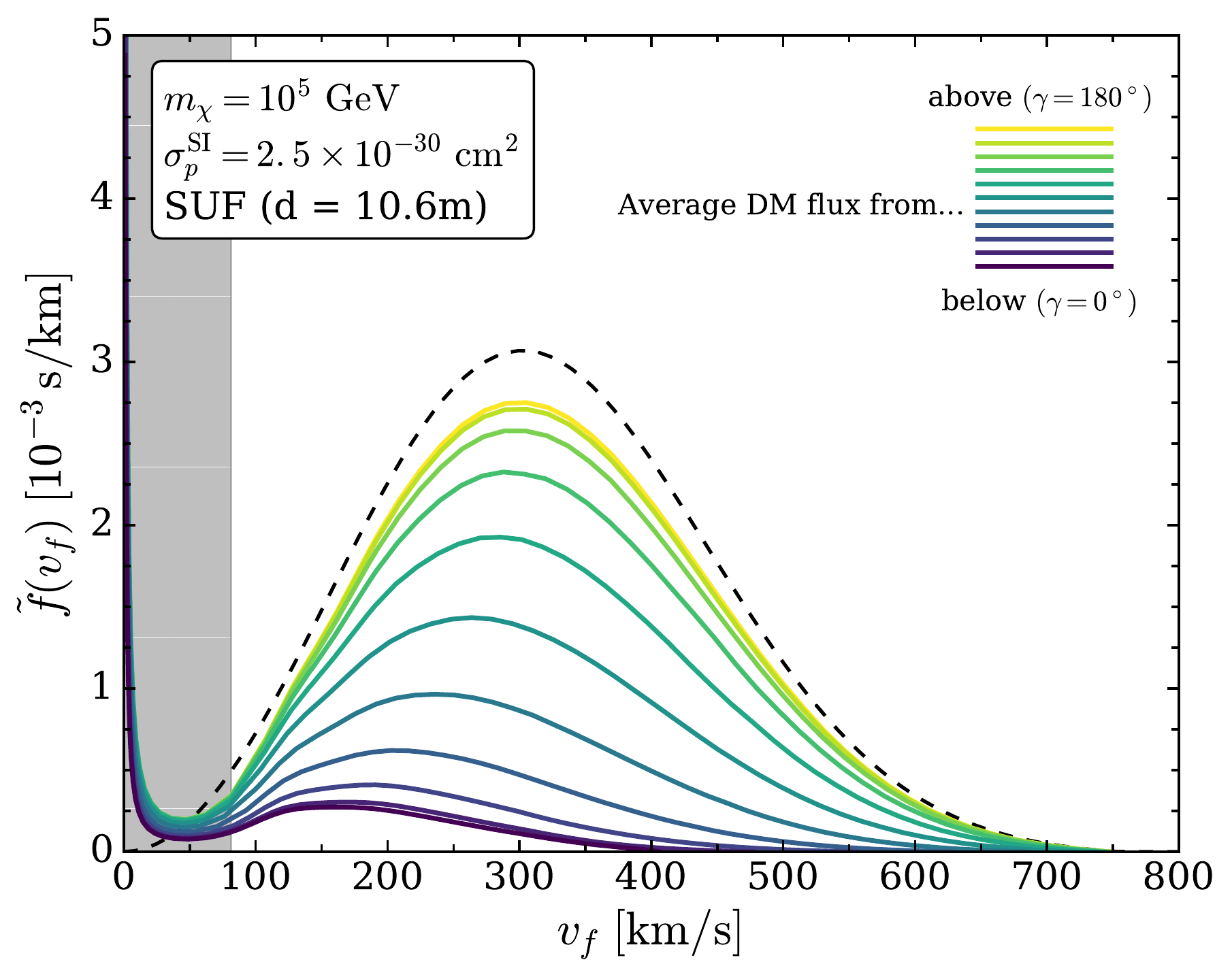}
\caption{\textbf{Final speed distribution at an underground detector at SUF.} Final speed distribution for DM particles of mass $10^5 \,\,\mathrm{GeV}$ at a detector at the Stanford Underground Facility (SUF) after propagation through the atmosphere, Earth and lead shielding. The dashed black lines show the unperturbed Maxwell-Boltzmann speed distribution. Speeds in the grey shaded region are typically too small to excite a nuclear recoil of energy 10 keV, the analysis threshold for CDMS-I. The speed distribution is shown as a function of the direction of the average DM flux (for fixed cross section). The different curves (from top to bottom) show results for equally spaced $\gamma$ values from $\gamma = 180^\circ$ (average flux from overhead) to $\gamma = 0^\circ$ (average flux from below).}
\label{fig:SpeedDists_gamma}
\end{figure}

\begin{figure*}[t]
\centering
\includegraphics[width=0.49\textwidth,]{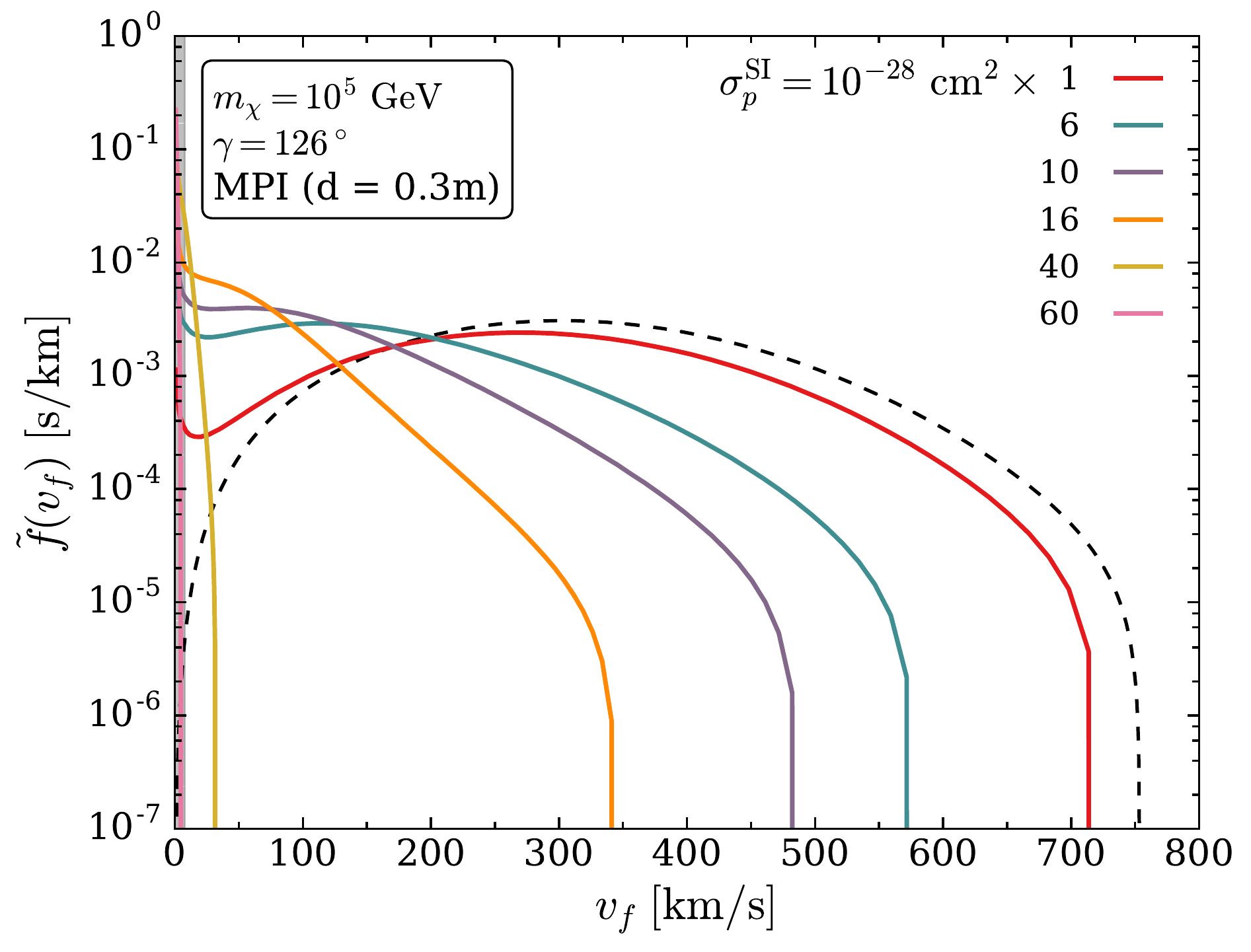}
\includegraphics[width=0.49\textwidth,]{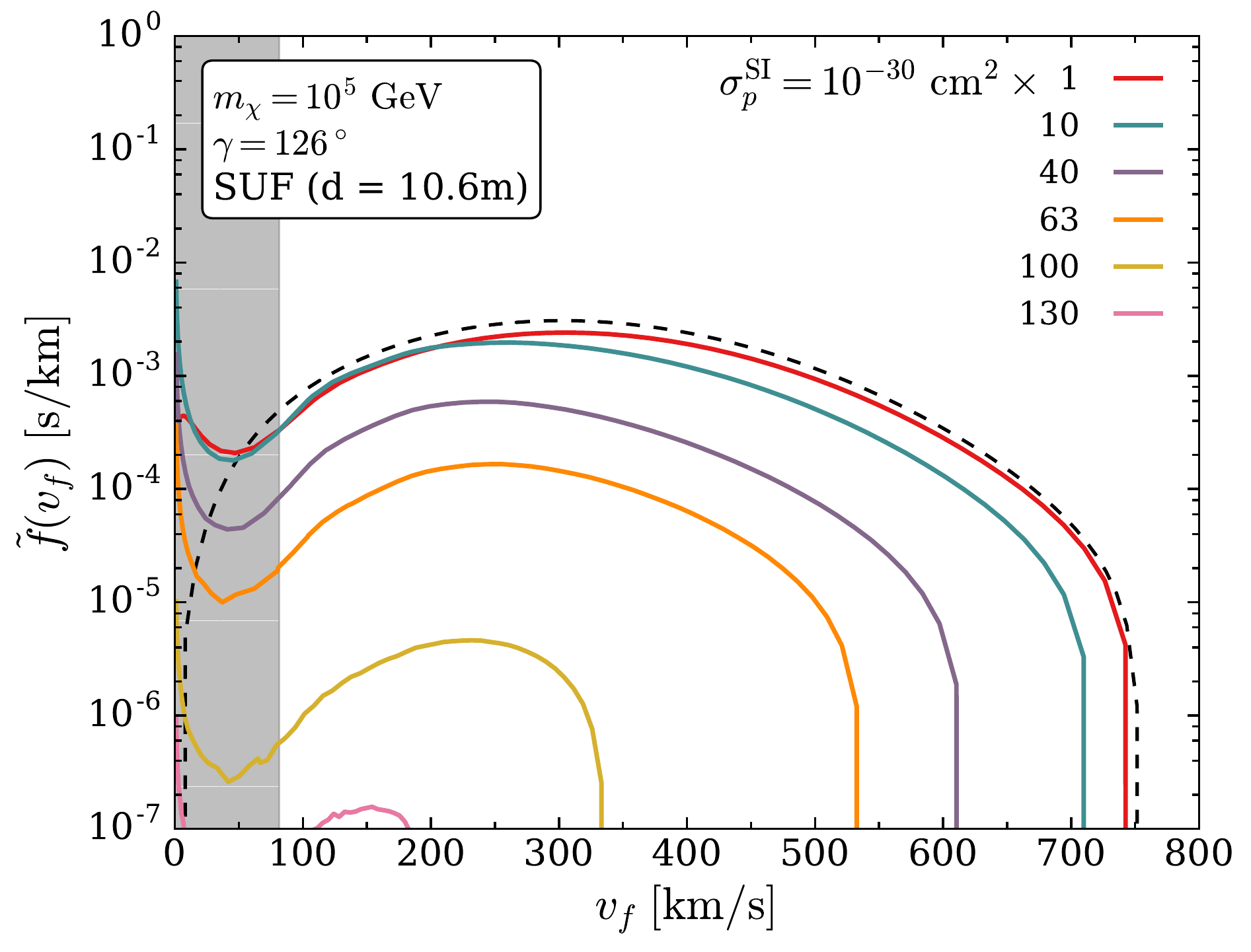}
\caption{\textbf{Final speed distribution at different detector locations for a range of DM-nucleon cross sections.} Final speed distribution for DM particles of mass $10^5 \,\,\mathrm{GeV}$ at a detector at the Max-Planck-Institute Munich (MPI, \textbf{left}) and Stanford Underground Facility (SUF, \textbf{right}) after propagation through the atmosphere, Earth and shielding. The dashed black lines show the unperturbed Maxwell-Boltzmann speed distribution. Speeds in the grey shaded region are (on average) too small to excite a nuclear recoil above threshold in the corresponding detector. Note the different cross section values in the two panels.}
\label{fig:SpeedDists_xsec}
\end{figure*}

Figure~\ref{fig:SpeedDists_gamma} shows the final DM speed distribution at a detector at SUF for a range of values of $\gamma$, the direction of the average DM flux. The attenuation effect becomes more pronounced as we go from $\gamma = 180^\circ$ to $\gamma = 0^\circ$. In the former case, DM particles have a shorter path  on average to reach the detector, while in the latter, the typical DM particle must cross the entire Earth to reach the detector. We note, however, that $\gamma$ denotes only the \textit{average} incoming DM direction; the DM particles have a distribution of incoming directions. This is why the $\gamma = 0^\circ$ case does not lead to complete attenuation: some particles still arrive at the detector from roughly overhead (though these typically have smaller speeds in the first place).

In Fig.~\ref{fig:SpeedDists_xsec}, we show the final speed distribution as a function of the interaction cross section for detectors at both MPI (left) and SUF (right), fixing the mean incoming DM direction to $\gamma = 126^\circ$ (a typical value for both detector sites). As expected, increasing the DM-nucleon cross section reduces the maximum speed of particles which arrive at the detector and increasingly populates the low-speed tail of the distribution. For a detector near the surface (left), the final speed distribution typically resembles a `compressed' version of the initial distribution. This is because the final speed scales roughly linearly with the initial speed as demonstrated in the left panel of Fig.~\ref{fig:VelocityTransfer}. The result is a roughly exponential cut off in the final speed distribution, with the maximum DM speed decreasing with increasing cross section.

In the case of SUF (right panel of Fig.~\ref{fig:SpeedDists_xsec}), the final speed distribution is not simply a rescaled Maxwell-Boltzmann. Instead, slower particles tend to be stopped almost completely (populating the low-speed tail) while faster particles experience comparatively less stopping (due to the coherence factor $C(m_\chi, v)$), leading to a bump at higher speeds. Because the speed distribution does not drop  as rapidly as for a detector at MPI, we expect that a large rate should be observable by a detector at SUF, so long as the maximum final speed lies above the threshold speed of the experiment (shown as a grey band). 



\section{Constraints}
\label{sec:constraints}

In order to determine the maximum cross section which can be probed by the CRESST 2017 surface run and the CDMS-I run at SUF, we must compare the observed number of recoil events with the predicted signal, given by Eq.~\eqref{eq:Nexp}, evaluated by performing a grid scan over $m_\chi$ and $\sigma_p^\mathrm{SI}$. Assuming (conservatively) that all observed events could be signal, we set a single-bin Poisson limit at the 90\% confidence level on the minimum value of the cross section. Given the density of the grid scan, we expect these lower limits to be accurate at the level of 2-3\%.

For CDMS-I, we consider an analysis window for nuclear recoil energies of $E_R \in [10, 100]$ keV and use an estimated Gaussian energy resolution of 2.4 keV \cite{Abrams:2002nb}. The Germanium detector modules at CDMS-I observed 27 veto anti-coincident events, consistent with single-scatter and multiple-scatter nuclear recoils. The total exposure  was $MT = 15.8\,\,\mathrm{kg}\,\, \mathrm{days}$, obtained over roughly 1 year in 1998-1999. In CDMS-I, the signal rate varies rapidly with cross section; the `bump' in the final speed distribution (shown in the right panel of Fig.~\ref{fig:SpeedDists_xsec}) means that the signal rate is very large as long as the maximum DM speed lies above the experimental threshold. For a DM mass of $10^5 \,\,\mathrm{GeV}$, we limit the cross section to be larger than $\sigma_p^\mathrm{SI} = 1.55 \times 10^{-28}\,\,\mathrm{cm}^2$. We have also checked that in this case, the total energy deposited by a single particle in a 1cm-thick Ge detector module (due to multiple-scattering) is $\mathcal{O}$(20 keV), meaning that such a particle should fall within the analysis window of the experiment and be in principle detectable.


For the CRESST 2017 surface run, using an $\mathrm{Al}_2\mathrm{O}_3$ target, we assume a total exposure of $4.6 \times 10^{-5}\,\,\mathrm{kg} \,\mathrm{days}$, an energy window of $E_R \in [19.7, \, 600] \,\mathrm{eV}$ and a Gaussian energy resolution of 3.74 eV at threshold. In this exposure, 511 events were observed in the region of interest. Because of the exponential cut off of the speed distribution in this case (see the left panel of Fig.~\ref{fig:SpeedDists_xsec}), the number of signal events varies more slowly with cross section. For a DM mass of $10^5 \,\,\mathrm{GeV}$, we limit the cross section to be larger than $\sigma_p^\mathrm{SI} = 4.5 \times 10^{-27}\,\,\mathrm{cm}^2$. As for CDMS-I, the limit for larger masses can be obtained by simply rescaling this cross section by a factor $m_\chi/(10^5\,\,\mathrm{GeV})$.

\begin{figure}[t]
\centering
\includegraphics[width=0.49\textwidth,]{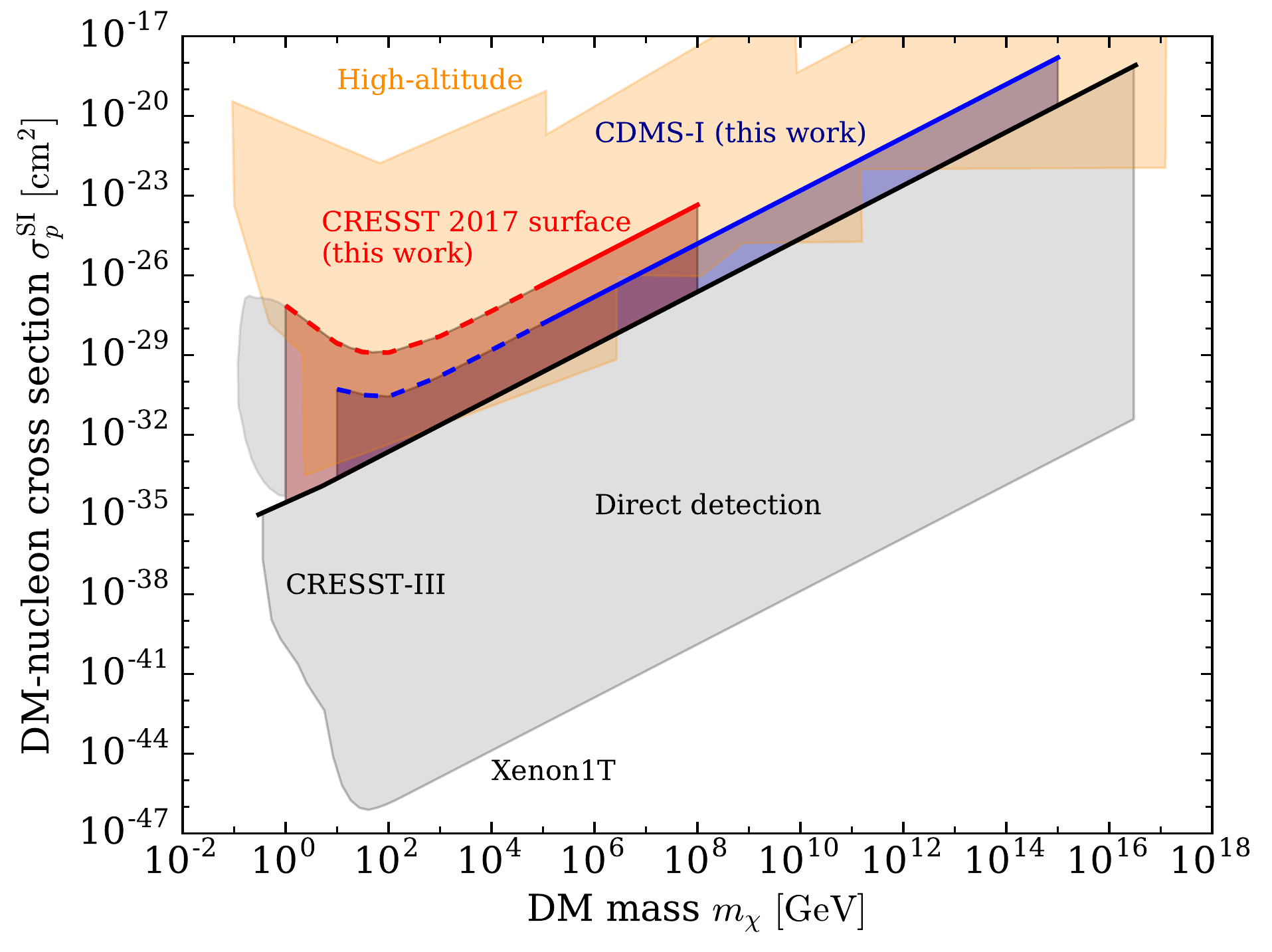}
\caption{\textbf{Constraints on strongly interacting Dark Matter from nuclear scattering search experiments.} The orange shaded region is excluded by high-altitude experiments such as RSS \cite{Rich1987}, XQC \cite{Zaharijas:2004jv,Erickcek:2007jv}, IMP 7/8 \cite{SnowdenIfft1990}, IMAX \cite{McGuire:1994pq} and SKYLAB \cite{Shirk1978} (collected in Ref.~\cite{Mack:2007xj}). The grey shaded regions are excluded by previous analyses of direct detection experiments. This region is bounded from below by constraints from Xenon1T \cite{Aprile:2017iyp} and from the left by CRESST-III \cite{Petricca:2017zdp} and the CRESST 2017 surface run \cite{Angloher:2017sxg,Davis:2017noy}. The solid black line denotes the previous cross section reach from Ref.~\cite{Albuquerque:2003ei} (reported in Ref.~\cite{Mack:2007xj}). The shaded blue area shows the additional region of parameter space excluded by the present reanalysis of the CDMS-I (SUF) limits \cite{Abusaidi:2000wg,Abrams:2002nb}, while the shaded red region shows the new limits from the CRESST 2017 surface run \cite{Angloher:2017sxg}. We also exclude the area below the dashed blue and red lines, though we note that all constraints on SIMPs in this mass range come with caveats, as discussed in the text.}
\label{fig:constraints}
\end{figure}

Figure~\ref{fig:constraints} shows the resulting constraints from CDMS-I in blue and from the CRESST 2017 surface run in red.  For CDMS-I, we extend the limit up to DM masses of $\mathcal{O}(10^{15})$ GeV. At such high masses, the number density of DM particles is very low and the rate of DM particles crossing a $10 \,\,\mathrm{cm}$-scale detector is $\mathcal{O}(1)$ per year. At larger masses than this, the limit from CDMS-I disappears, as we would not expect any substantial number of DM particles to impinge on the detector over the course of the 1 year exposure. In the case of the CRESST 2017 surface run, we truncate the limit at a mass of $\mathcal{O}(10^{8})$ GeV. For a $5 \,\,\mathrm{mm}$-scale crystal exposed over 2.27 hours, the number of $10^{8}$ GeV DM particles crossing the detector is $\mathcal{O}(200)$ and above this mass, the flux of DM particles is too small to explain the $\mathcal{O}(500)$ events in the CRESST 2017 surface run dataset.\footnote{We note also that the right hand limit of the grey shaded region in Fig.~\ref{fig:constraints} should not be taken as exact. While Xenon1T has a large enough area and exposure to probe up to DM masses  of around $10^{17}$ GeV, more detailed calculations (including stopping effects) are required to estimate where the limit should be truncated.}

 We also plot the limit obtained using our procedure in the mass range $1$ GeV to $10^5$ GeV. For DM particles heavier than around $50\,\mathrm{GeV}$ (the mass of an iron nucleus), the assumptions we have made here should not be too strongly violated. In particular, as long as $m_\chi > m_i$ the deflection of DM particles will be predominantly in the forward direction and so, with a large number of scatters, the typical deflection should not be too large. 

Instead, for light DM the limits should be treated with caution because the deflection of DM particles in a single scatter can become large and therefore the assumption of straight-line trajectories is violated. We note, however, that while a number of particles are deflected \textit{away} from the detector, some of these should be replaced by those particles which are deflected \textit{towards} the detector (c.f.~Ref.~\cite{Kavanagh:2016pyr}). For isotropic deflection (in the limit of low DM masses), these two effects should balance. The dashed bounds plotted in Fig.~\ref{fig:constraints} therefore correspond to such a scenario, where the net DM flux at the detector is unchanged and the typical path length of particles is not substantially increased by the deflections.
As discussed in Refs.~\cite{Mahdawi:2017cxz,Mahdawi:2017utm}, those particles reaching the detector (and contributing predominantly to the limit) may not correspond to the \textit{average} particles but to those which scattered less often than expected, with a resulting shorter path length than average. In the end, dedicated Monte Carlo simulations (using codes such as \textsc{DaMaSCUS} \cite{Emken2017a, Emken:2017qmp}) will be required to obtain a precise limit in this region.

The results of Fig.~\ref{fig:constraints} indicate that constraints from the CDMS-I run at the Stanford Underground Facility extend up to larger cross sections than previously thought, by as much as 2 orders of magnitude over a wide range of DM masses. In addition, the constraints from CRESST 2017 surface run extend the limits by an additional factor of $\sim50$ for DM masses below $\sim 10^{8}\,\,\mathrm{GeV}$. In particular, these revised constraints now close a number of gaps in the parameter space which persisted between underground experiments and high-altitude experiments in the mass range between $10^7$ and $10^{13}$ GeV. Though these parameter windows are also disfavoured by limits on anomalous heating in the Earth \cite{Mack:2007xj,Mack2013} and neutrino telescope constraints \cite{Albuquerque:2010bt}, we have demonstrated that they are also ruled out independently by dedicated direct searches for DM-nuclear recoils.

Where does the improvement in the CDMS-I constraints come from, compared to the previous analysis of Ref.~\cite{Albuquerque:2003ei}? The first contribution comes from accounting for the form factor suppression of the scattering rate, encoded in the coherence factor $C(m_\chi, v)$ in Eq.~\eqref{eq:stopping}. The impact of form factor suppression was demonstrated in Fig.~\ref{fig:VelocityTransfer}, where we observed that fast-moving particles are slowed less than would be expected from scattering off point-like nuclei in the Earth and shielding. While initially slow-moving particles are effectively stopped,  a fast-moving population is preserved and can reach the detector.

The second contribution comes from considering the full (lab-frame anisotropic) 3-dimensional distribution of DM velocities. In Fig.~\ref{fig:gamma}, we showed that at SUF, the flux of DM particles appears to come mostly from above. This means that the typical DM particle must travel a much shorter path length to reach the detector than would be expected from an isotropic DM flux (as assumed in Ref.~\cite{Albuquerque:2003ei}). The impact of this is clear in the left panel of Fig.~\ref{fig:SpeedDists_xsec}: particles impinging from above experience a much smaller stopping effect and arrive at the detector with a larger speed. The overall effect is that DM particles with much larger cross sections than previously thought can arrive at the detector with appreciable speed, increasing the maximum cross section reach of the experiments.

How robust are these constraints? Here we have assumed that the free, asymptotic velocity distribution matches the Standard Halo Model, which is typically assumed for WIMP Dark Matter. Of course, strong interactions could alter this distribution (see e.g.~Ref.~\cite{Fan:2013yva}) before the particles ever reach the Earth. In addition, there are also uncertainties on the WIMP velocity distribution from N-body simulations \cite{Bozorgnia:2017brl}. We could of course study specific velocity distributions using the methods presented here, but unless these lead to a substantial change in the maximum velocity of SIMPs arriving at Earth, we do not expect the limits to change by a large amount. Indeed, the largest effect would perhaps come from uncertainties in the Galactic escape speed itself, $v_\mathrm{esc} \in [492, 586] \,\,\mathrm{km/s}$ (90\% confidence) \cite{Piffl:2013mla}. A larger escape speed would increase the maximum cross section which can be probed by underground direct detection experiments. As can be seen in Fig.~\ref{fig:SpeedDists_xsec}, increasing the cross section by a few tens of percent can reduce the maximum velocity of DM particles arriving at the detector by over $100 \,\,\mathrm{km/s}$. We would therefore expect uncertainties in $v_\mathrm{esc}$ to contribute at most an $\mathcal{O}(30\%)$ correction to the bounds we have derived here.

As we argued in the introduction, the  angular deflection of DM particles (per scatter) is $\delta \theta \sim m_N/m_\chi$. As we move to larger DM masses and cross sections, the deflection per scatter decreases while the number of scatterings increases. This means that more and more DM particles will follow trajectories close to straight lines and the average energy loss (Eq.~\eqref{eq:energyloss}) will become an increasingly accurate description of the entire population of particles. Of course, DM particles which undergo fewer scatters than the average will typically arrive at the detector with a larger speed, meaning that there could still be a (highly suppressed) high speed tail of particles not captured by our description. Refs.~\cite{Mahdawi:2017cxz,Mahdawi:2017utm} suggest an importance sampling method to capture these effects, though the relevant code has not yet been released. In any case, given the large number of scattering events we consider here, along with the rapid change in the CDMS-I event rate with cross section, we do not expect this high speed tail to significantly alter our results.

\section{Conclusions}

In this work, we have presented limits on strongly-interacting massive particle (SIMP) Dark Matter from direct detection experiments. We take into account the stopping effect of the atmosphere, Earth and detector shielding using a nuclear-stopping formalism. The DM particles are assumed to scatter continuously and travel along straight line trajectories, an approach we argue is valid for very heavy DM ($m_\chi \gtrsim 10^5 \,\,\mathrm{GeV}$). We account for the full 3-dimensional velocity distribution of the incoming DM, as well as form factor suppression of the DM-nucleus scattering, which leads to smaller energy losses for high speed particles than previously assumed. 

\begin{figure}[t]
\centering
\includegraphics[width=0.49\textwidth,]{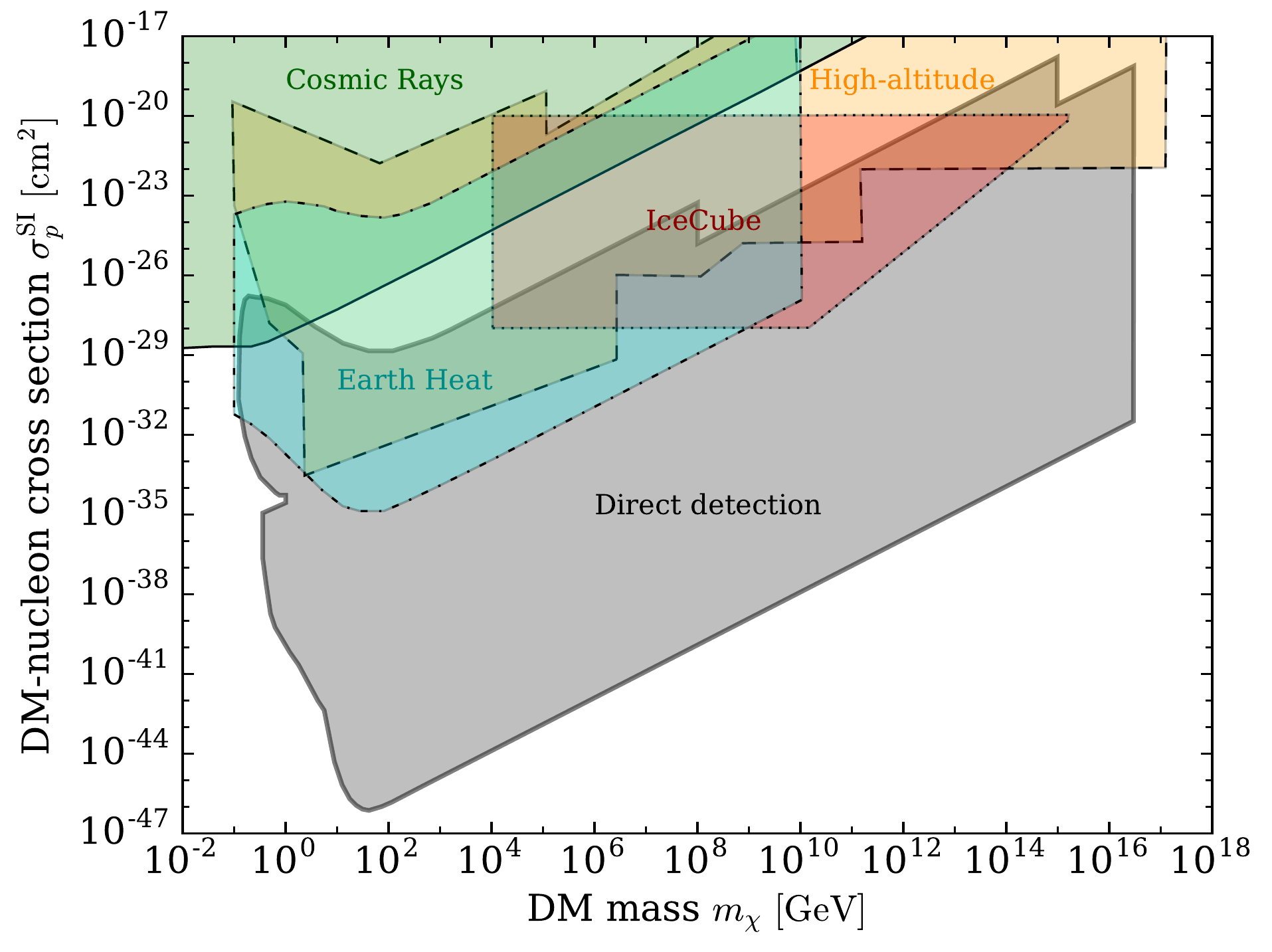}
\caption{\textbf{Summary of constraints on strongly-interacting Dark Matter.} The solid grey region shows constraints from surface and underground direct detection experiments, including the updated limits derived in this work. The orange region (dashed) is excluded by high-altitude experiments, as in Fig.~\ref{fig:constraints}. Complementary constraints from IceCube (red, dotted) \cite{Albuquerque:2010bt}, Earth's heat flux (cyan, dot-dashed) \cite{Mack:2007xj,Mack2013} and DM-cosmic ray interactions (green, solid) \cite{Cyburt2002} are also shown. Note that these complementary constraints are model-dependent and typically require further assumptions about Dark Matter, beyond its nucleon scattering cross section (e.g.~self-annihilation into Standard Model states). Large areas of the SIMP Dark Matter parameter space are ruled out by at least two constraints, both direct searches for DM-nuclear scattering and indirect constraints.}
\label{fig:constraints_all}
\end{figure}

Recent results from the CRESST 2017 surface run \cite{Angloher:2017sxg} allow us probe up to roughly 5000 times larger cross sections than was previously possible. For masses larger than about $10^{8}\,\,\mathrm{GeV}$, the rate of DM particles crossing the detector during the CRESST 2017 surface run is small and a detector with a larger exposure time is required. Our reanalysis of the CDMS-I experiment \cite{Abusaidi:2000wg,Abrams:2002nb}, operated at a depth of around 10m, extends previous limits \cite{Albuquerque:2003ei} by two orders of magnitude in cross section, up to DM masses $\mathcal{O}(10^{15})\,\,\mathrm{GeV}$. This highlights the importance of careful modelling when studying the Earth-scattering of Dark Matter. With this in mind, we note that while we also present limits on DM masses down to 1 GeV, in that case our method is not strictly valid, and such limits must be confirmed with dedicated Monte Carlo simulations. 

These updated constraints close a number of windows in the parameter space in the mass range $10^6$ GeV to $10^{13}$ GeV, between air-borne and high-altitude experiments on the one hand and ground-based and underground experiments on the other. For super-heavy DM produced gravitationally at the end of inflation \cite{Hui:1998dc,Allahverdi:2002nb}, the DM mass is typically around the inflaton mass scale, which lies precisely in this range \cite{Brout:2002ec}. These constraints, which come solely from direct searches for DM-nucleus scattering, are complemented by independent indirect constraints from DM-cosmic ray interactions \cite{Cyburt2002}, from IceCube searches for the annihilation products of SIMPs captured in the Sun \cite{Albuquerque:2010bt} and from limits on the heat flux from the Earth (which would be affected by SIMP capture and annihilation) \cite{Mack:2007xj,Mack2013}. We emphasise however that several of these indirect constraints rely on further assumptions about the Dark Matter particle beyond its DM-nucleon scattering properties (such as its self-annihilation into Standard Model particles).  

The current constraints are summarised in Fig.~\ref{fig:constraints_all}. With the updated constraints presented here, SIMP Dark Matter is now ruled out in the approximate range $\sigma_p^{\mathrm{SI}}/m_\chi \in [10^{-37},\, 10^{-25}] \,\,\mathrm{cm}^2\,/\mathrm{GeV}$, in many cases using multiple independent probes, relying on different sets of assumptions.



\acknowledgements

The author thanks Laura Baudis, Jonathan Davis and Timon Emken for helpful discussions and comments on an early version of this paper. The author also thanks Ciaran O'Hare for providing useful code for calculating the lab velocity as a function of time. The author gratefully acknowledges support from the Netherlands Organisation for Scientific Research (NWO) and from the European Research Council ({\sc Erc}) under the EU Seventh Framework Programme (FP7/2007-2013)/{\sc Erc} Starting Grant (agreement n.\ 278234 --- `{\sc NewDark}' project).
This research was also supported by the Munich Institute for Astro- and Particle Physics (MIAPP) of the DFG cluster of excellence ``Origin and Structure of the Universe".
Part of this work was carried out on the Dutch national e-infrastructure with the support of SURF Cooperative.

\appendix

\section{Coordinate system}
\label{app:coordinates}
\begin{figure}[t!]
\centering
\includegraphics[width=0.30\textwidth,]{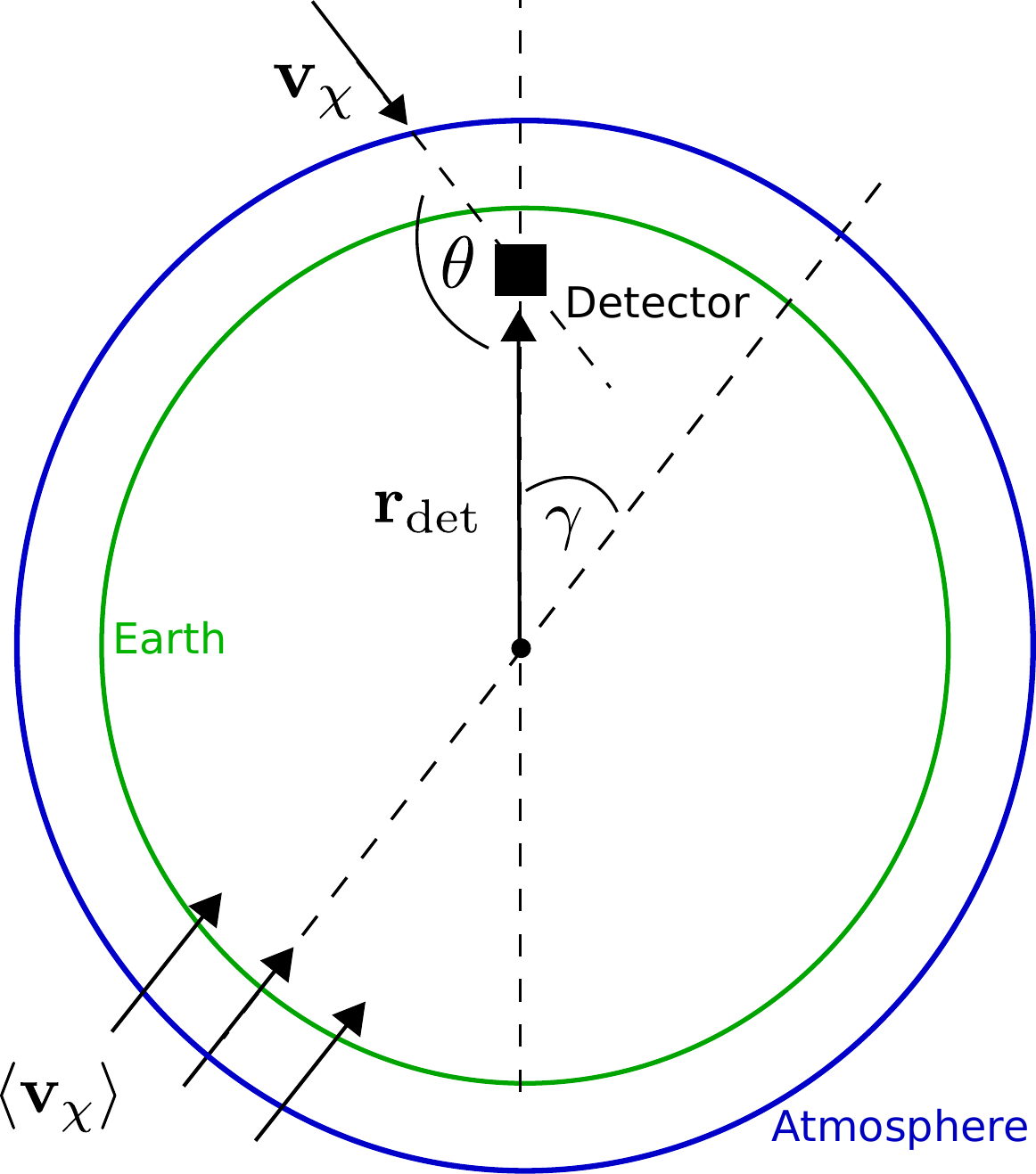}
\caption{\textbf{Coordinate system for Earth scattering.} Illustration of the coordinate system used to calculate the impact of DM scattering in the Earth. The incoming direction of a single DM particle towards the detector is denoted by $\mathbf{v}$, while the direction of the mean DM flux is denoted $\langle \mathbf{v}_\chi\rangle = -\mathbf{v}_\mathrm{lab}(t)$. In the associated text, we give expressions for the DM path length to the detector and the velocity distribution in this coordinate system.}
\label{fig:earth}
\end{figure}

Here, we give explicit expressions for the DM path length and velocity distribution in terms of the angular coordinates specifying the DM trajectory and the detector position (with respect to the mean DM flux direction). This coordinate system is illustrated in Fig.~\ref{fig:earth}. We denote the detector position as $\mathbf{r}_\mathrm{det}$ (as measured from the centre of the Earth) and we assume that the detector is at a vertical depth $d$ from the surface of the Earth. We denote the Earth radius as $R_E \approx 6371\,\mathrm{km}$ and the vertical height of the atmosphere as $h_A \approx 80 \,\mathrm{km}$. With these definitions, we have $|\mathbf{r}_\mathrm{det}| = r_\mathrm{det}= R_E - d$. 

We assume that a specific DM particle arrives at the detector in direction $\hat{\mathbf{v}}$, which makes an angle $\theta$ with the vertical.\footnote{Note that $\theta = 0$ corresponds to an \textit{upward}-going DM particle.} The straight-line path length $\ell$ from the top of the atmosphere to the detector for such a particle is given by:
\begin{equation}
\ell = r_\mathrm{det} \cos\theta + \sqrt{(R_E + h_A)^2 - (r_\mathrm{det} \sin\theta)^2 }\,.
\end{equation}

In order to correctly compute the scattering rate for DM particles, we must account for the radial dependence of the nuclear density in the Earth and atmosphere. For a DM particle which has travelled a distance $D$ along its trajectory, as measured from the top of the atmosphere, the radial distance from the Earth's centre is:
\begin{equation}
r = \sqrt{(R_E + h_A)^2 + D^2 + 2D(r_\mathrm{det}\cos\theta - \ell)}\,.
\end{equation}
With this, we can calculate the final speed $v_f$ from the initial speed $v_i$ (at the top of the atmosphere), along a given trajectory, using
\begin{equation}
\label{eq:vf}
v_f = v_i + \int_0^{\ell} \frac{\mathrm{d}v}{\mathrm{d}D}(v, r) \,\mathrm{d}D\,.
\end{equation}
We emphasise that  $\mathrm{d}v/\mathrm{d}D$ (given explicitly in Eq.~\eqref{eq:stopping}) depends on both $r$ and $v$ and so Eq.~\eqref{eq:vf} must be solved numerically using an ODE solver.

The DM velocity distribution is given in vectorial form in Eq.~\eqref{eq:veldist}. Here, we rewrite this in terms of angular coordinates. We describe the incoming DM direction using $\theta$, the polar angle measured from the downward vertical at the detector, and $\phi$, the azimuthal angle (which is suppressed in Fig.~\ref{fig:earth}). With these definitions, we can write:
\begin{equation}
|\mathbf{v} - \langle \mathbf{v}_\chi\rangle|^2 = v^2 - 2vv_\mathrm{lab} \cos\delta + v_\mathrm{lab}^2\,,
\end{equation}
where 
\begin{equation}
\cos\delta = \sin\gamma \sin\theta \cos\phi + \cos\gamma \cos\theta\,.
\end{equation}
In calculating the velocity distribution, we fix the magnitude of $v_\mathrm{lab} = 220 \,\mathrm{km/s}$, with the time dependence arising through the angle $\gamma = \gamma(t)$:
\begin{equation}
\gamma = \cos^{-1}\left(\langle \hat{\mathbf{v}}_\chi\rangle \cdot \hat{\mathbf{r}}_\mathrm{det}\right)\,.
\end{equation}

We finally note that path length of the DM does not depend on the azimuthal angle $\phi$, meaning that the final velocity at the detector depends only on the initial velocity and the polar angle, $v_f = v_f(v_i, \cos\theta)$. This means that in evaluating the final speed distribution at the detector (see Sec.~\ref{sec:VelDist}),
\begin{align}
\begin{split}
\tilde{f}(v_f,\gamma) = \oint f(\mathbf{v}_i,\gamma) \,v_i^2 \dbd{v_i}{v_f} \, \mathrm{d}^2\hat{\mathbf{v}}\,,
\end{split}
\end{align}
the two angular integrations can be performed separately. In terms of the angular coordinates, the final speed distribution is then,
\begin{align}
\begin{split}
\label{eq:fvf}
\tilde{f}(v_f,\gamma) = \int_{-1}^{1} f(v_i, \cos\theta, \gamma)\, v_i^2 \dbd{v_i}{v_f} \,\mathrm{d}\cos\theta\,,
\end{split}
\end{align}
where we define
\begin{align}
\begin{split}
\label{eq:fcostheta}
f(v_i, \cos\theta, \gamma) = \int_{0}^{2\pi} f(v_i, \cos\theta,\phi,\gamma)\,\mathrm{d}\phi\,.
\end{split}
\end{align}
The values of $f(v_i, \cos\theta, \gamma)$ are easily tabulated as a function of $v_i$ and $\cos\theta \cos\gamma$, meaning that only a single integration over $\theta$ is required (at a given value of $v_f$ and $\gamma$) in order to evaluate the final velocity distribution (Eq.~\refeq{eq:fvf}).

\bibliography{WIMPzillas.bib}

\begin{thebibliography}{71}%
\makeatletter
\providecommand \@ifxundefined [1]{%
 \@ifx{#1\undefined}
}%
\providecommand \@ifnum [1]{%
 \ifnum #1\expandafter \@firstoftwo
 \else \expandafter \@secondoftwo
 \fi
}%
\providecommand \@ifx [1]{%
 \ifx #1\expandafter \@firstoftwo
 \else \expandafter \@secondoftwo
 \fi
}%
\providecommand \natexlab [1]{#1}%
\providecommand \enquote  [1]{``#1''}%
\providecommand \bibnamefont  [1]{#1}%
\providecommand \bibfnamefont [1]{#1}%
\providecommand \citenamefont [1]{#1}%
\providecommand \href@noop [0]{\@secondoftwo}%
\providecommand \href [0]{\begingroup \@sanitize@url \@href}%
\providecommand \@href[1]{\@@startlink{#1}\@@href}%
\providecommand \@@href[1]{\endgroup#1\@@endlink}%
\providecommand \@sanitize@url [0]{\catcode `\\12\catcode `\$12\catcode
  `\&12\catcode `\#12\catcode `\^12\catcode `\_12\catcode `\%12\relax}%
\providecommand \@@startlink[1]{}%
\providecommand \@@endlink[0]{}%
\providecommand \url  [0]{\begingroup\@sanitize@url \@url }%
\providecommand \@url [1]{\endgroup\@href {#1}{\urlprefix }}%
\providecommand \urlprefix  [0]{URL }%
\providecommand \Eprint [0]{\href }%
\providecommand \doibase [0]{http://dx.doi.org/}%
\providecommand \selectlanguage [0]{\@gobble}%
\providecommand \bibinfo  [0]{\@secondoftwo}%
\providecommand \bibfield  [0]{\@secondoftwo}%
\providecommand \translation [1]{[#1]}%
\providecommand \BibitemOpen [0]{}%
\providecommand \bibitemStop [0]{}%
\providecommand \bibitemNoStop [0]{.\EOS\space}%
\providecommand \EOS [0]{\spacefactor3000\relax}%
\providecommand \BibitemShut  [1]{\csname bibitem#1\endcsname}%
\let\auto@bib@innerbib\@empty
\bibitem [{\citenamefont {Goodman}\ and\ \citenamefont
  {Witten}(1985)}]{Goodman:1984dc}%
  \BibitemOpen
  \bibfield  {author} {\bibinfo {author} {\bibfnamefont {M.~W.}\ \bibnamefont
  {Goodman}}\ and\ \bibinfo {author} {\bibfnamefont {E.}~\bibnamefont
  {Witten}},\ }\href {\doibase 10.1103/PhysRevD.31.3059} {\bibfield  {journal}
  {\bibinfo  {journal} {Phys. Rev.}\ }\textbf {\bibinfo {volume} {D31}},\
  \bibinfo {pages} {3059} (\bibinfo {year} {1985})}\BibitemShut {NoStop}%
\bibitem [{\citenamefont {Drukier}\ \emph {et~al.}(1986)\citenamefont
  {Drukier}, \citenamefont {Freese},\ and\ \citenamefont
  {Spergel}}]{Drukier:1986tm}%
  \BibitemOpen
  \bibfield  {author} {\bibinfo {author} {\bibfnamefont {A.~K.}\ \bibnamefont
  {Drukier}}, \bibinfo {author} {\bibfnamefont {K.}~\bibnamefont {Freese}}, \
  and\ \bibinfo {author} {\bibfnamefont {D.~N.}\ \bibnamefont {Spergel}},\
  }\href {\doibase 10.1103/PhysRevD.33.3495} {\bibfield  {journal} {\bibinfo
  {journal} {Phys. Rev.}\ }\textbf {\bibinfo {volume} {D33}},\ \bibinfo {pages}
  {3495} (\bibinfo {year} {1986})}\BibitemShut {NoStop}%
\bibitem [{\citenamefont {Aprile}\ \emph {et~al.}(2017)\citenamefont {Aprile}
  \emph {et~al.}}]{Aprile:2017iyp}%
  \BibitemOpen
  \bibfield  {author} {\bibinfo {author} {\bibfnamefont {E.}~\bibnamefont
  {Aprile}} \emph {et~al.} (\bibinfo {collaboration} {XENON}),\ }\href
  {\doibase 10.1103/PhysRevLett.119.181301} {\bibfield  {journal} {\bibinfo
  {journal} {Phys. Rev. Lett.}\ }\textbf {\bibinfo {volume} {119}},\ \bibinfo
  {pages} {181301} (\bibinfo {year} {2017})},\ \Eprint
  {http://arxiv.org/abs/1705.06655} {arXiv:1705.06655 [astro-ph.CO]}
  \BibitemShut {NoStop}%
\bibitem [{\citenamefont {Cui}\ \emph {et~al.}(2017)\citenamefont {Cui} \emph
  {et~al.}}]{Cui:2017nnn}%
  \BibitemOpen
  \bibfield  {author} {\bibinfo {author} {\bibfnamefont {X.}~\bibnamefont
  {Cui}} \emph {et~al.} (\bibinfo {collaboration} {PandaX-II}),\ }\href
  {\doibase 10.1103/PhysRevLett.119.181302} {\bibfield  {journal} {\bibinfo
  {journal} {Phys. Rev. Lett.}\ }\textbf {\bibinfo {volume} {119}},\ \bibinfo
  {pages} {181302} (\bibinfo {year} {2017})},\ \Eprint
  {http://arxiv.org/abs/1708.06917} {arXiv:1708.06917 [astro-ph.CO]}
  \BibitemShut {NoStop}%
\bibitem [{\citenamefont {Petricca}\ \emph {et~al.}(2017)\citenamefont
  {Petricca} \emph {et~al.}}]{Petricca:2017zdp}%
  \BibitemOpen
  \bibfield  {author} {\bibinfo {author} {\bibfnamefont {F.}~\bibnamefont
  {Petricca}} \emph {et~al.} (\bibinfo {collaboration} {CRESST}),\ }in\ \href
  {https://inspirehep.net/record/1637341/files/arXiv:1711.07692.pdf} {\emph
  {\bibinfo {booktitle} {{15th International Conference on Topics in
  Astroparticle and Underground Physics (TAUP 2017) Sudbury, Ontario, Canada,
  July 24-28, 2017}}}}\ (\bibinfo {year} {2017})\ \Eprint
  {http://arxiv.org/abs/1711.07692} {arXiv:1711.07692 [astro-ph.CO]}
  \BibitemShut {NoStop}%
\bibitem [{\citenamefont {Angloher}\ \emph {et~al.}(2017)\citenamefont
  {Angloher} \emph {et~al.}}]{Angloher:2017sxg}%
  \BibitemOpen
  \bibfield  {author} {\bibinfo {author} {\bibfnamefont {G.}~\bibnamefont
  {Angloher}} \emph {et~al.},\ }\href {\doibase 10.1140/epjc/s10052-017-5223-9}
  {\bibfield  {journal} {\bibinfo  {journal} {The European Physical Journal C}\
  }\textbf {\bibinfo {volume} {77}} (\bibinfo {year} {2017}),\
  10.1140/epjc/s10052-017-5223-9},\ \Eprint {http://arxiv.org/abs/1707.06749}
  {arXiv:1707.06749 [astro-ph.CO]} \BibitemShut {NoStop}%
\bibitem [{\citenamefont {Essig}\ \emph {et~al.}(2017)\citenamefont {Essig},
  \citenamefont {Volansky},\ and\ \citenamefont {Yu}}]{Essig:2017kqs}%
  \BibitemOpen
  \bibfield  {author} {\bibinfo {author} {\bibfnamefont {R.}~\bibnamefont
  {Essig}}, \bibinfo {author} {\bibfnamefont {T.}~\bibnamefont {Volansky}}, \
  and\ \bibinfo {author} {\bibfnamefont {T.-T.}\ \bibnamefont {Yu}},\ }\href
  {\doibase 10.1103/PhysRevD.96.043017} {\bibfield  {journal} {\bibinfo
  {journal} {Phys. Rev.}\ }\textbf {\bibinfo {volume} {D96}},\ \bibinfo {pages}
  {043017} (\bibinfo {year} {2017})},\ \Eprint
  {http://arxiv.org/abs/1703.00910} {arXiv:1703.00910 [hep-ph]} \BibitemShut
  {NoStop}%
\bibitem [{\citenamefont {Knapen}\ \emph {et~al.}(2017)\citenamefont {Knapen},
  \citenamefont {Lin},\ and\ \citenamefont {Zurek}}]{Knapen:2017xzo}%
  \BibitemOpen
  \bibfield  {author} {\bibinfo {author} {\bibfnamefont {S.}~\bibnamefont
  {Knapen}}, \bibinfo {author} {\bibfnamefont {T.}~\bibnamefont {Lin}}, \ and\
  \bibinfo {author} {\bibfnamefont {K.~M.}\ \bibnamefont {Zurek}},\ }\href@noop
  {} {\  (\bibinfo {year} {2017})},\ \Eprint {http://arxiv.org/abs/1709.07882}
  {arXiv:1709.07882 [hep-ph]} \BibitemShut {NoStop}%
\bibitem [{\citenamefont {An}\ \emph {et~al.}(2015)\citenamefont {An},
  \citenamefont {Pospelov}, \citenamefont {Pradler},\ and\ \citenamefont
  {Ritz}}]{An:2014twa}%
  \BibitemOpen
  \bibfield  {author} {\bibinfo {author} {\bibfnamefont {H.}~\bibnamefont
  {An}}, \bibinfo {author} {\bibfnamefont {M.}~\bibnamefont {Pospelov}},
  \bibinfo {author} {\bibfnamefont {J.}~\bibnamefont {Pradler}}, \ and\
  \bibinfo {author} {\bibfnamefont {A.}~\bibnamefont {Ritz}},\ }\href {\doibase
  10.1016/j.physletb.2015.06.018} {\bibfield  {journal} {\bibinfo  {journal}
  {Phys. Lett.}\ }\textbf {\bibinfo {volume} {B747}},\ \bibinfo {pages} {331}
  (\bibinfo {year} {2015})},\ \Eprint {http://arxiv.org/abs/1412.8378}
  {arXiv:1412.8378 [hep-ph]} \BibitemShut {NoStop}%
\bibitem [{\citenamefont {Kusenko}\ and\ \citenamefont
  {Shaposhnikov}(1998)}]{Kusenko:1997si}%
  \BibitemOpen
  \bibfield  {author} {\bibinfo {author} {\bibfnamefont {A.}~\bibnamefont
  {Kusenko}}\ and\ \bibinfo {author} {\bibfnamefont {M.~E.}\ \bibnamefont
  {Shaposhnikov}},\ }\href {\doibase 10.1016/S0370-2693(97)01375-0} {\bibfield
  {journal} {\bibinfo  {journal} {Phys. Lett.}\ }\textbf {\bibinfo {volume}
  {B418}},\ \bibinfo {pages} {46} (\bibinfo {year} {1998})},\ \Eprint
  {http://arxiv.org/abs/hep-ph/9709492} {arXiv:hep-ph/9709492 [hep-ph]}
  \BibitemShut {NoStop}%
\bibitem [{\citenamefont {Chung}\ \emph {et~al.}(1998)\citenamefont {Chung},
  \citenamefont {Kolb},\ and\ \citenamefont {Riotto}}]{Chung:1998ua}%
  \BibitemOpen
  \bibfield  {author} {\bibinfo {author} {\bibfnamefont {D.~J.~H.}\
  \bibnamefont {Chung}}, \bibinfo {author} {\bibfnamefont {E.~W.}\ \bibnamefont
  {Kolb}}, \ and\ \bibinfo {author} {\bibfnamefont {A.}~\bibnamefont
  {Riotto}},\ }\href {\doibase 10.1103/PhysRevLett.81.4048} {\bibfield
  {journal} {\bibinfo  {journal} {Phys. Rev. Lett.}\ }\textbf {\bibinfo
  {volume} {81}},\ \bibinfo {pages} {4048} (\bibinfo {year} {1998})},\ \Eprint
  {http://arxiv.org/abs/hep-ph/9805473} {arXiv:hep-ph/9805473 [hep-ph]}
  \BibitemShut {NoStop}%
\bibitem [{\citenamefont {Kolb}\ \emph {et~al.}(2007)\citenamefont {Kolb},
  \citenamefont {Starobinsky},\ and\ \citenamefont {Tkachev}}]{Kolb:2007vd}%
  \BibitemOpen
  \bibfield  {author} {\bibinfo {author} {\bibfnamefont {E.~W.}\ \bibnamefont
  {Kolb}}, \bibinfo {author} {\bibfnamefont {A.~A.}\ \bibnamefont
  {Starobinsky}}, \ and\ \bibinfo {author} {\bibfnamefont {I.~I.}\ \bibnamefont
  {Tkachev}},\ }\href {\doibase 10.1088/1475-7516/2007/07/005} {\bibfield
  {journal} {\bibinfo  {journal} {JCAP}\ }\textbf {\bibinfo {volume} {0707}},\
  \bibinfo {pages} {005} (\bibinfo {year} {2007})},\ \Eprint
  {http://arxiv.org/abs/hep-th/0702143} {arXiv:hep-th/0702143 [hep-th]}
  \BibitemShut {NoStop}%
\bibitem [{\citenamefont {Kolb}\ and\ \citenamefont
  {Long}(2017)}]{Kolb:2017jvz}%
  \BibitemOpen
  \bibfield  {author} {\bibinfo {author} {\bibfnamefont {E.~W.}\ \bibnamefont
  {Kolb}}\ and\ \bibinfo {author} {\bibfnamefont {A.~J.}\ \bibnamefont
  {Long}},\ }\href {\doibase 10.1103/PhysRevD.96.103540} {\bibfield  {journal}
  {\bibinfo  {journal} {Phys. Rev.}\ }\textbf {\bibinfo {volume} {D96}},\
  \bibinfo {pages} {103540} (\bibinfo {year} {2017})},\ \Eprint
  {http://arxiv.org/abs/1708.04293} {arXiv:1708.04293 [astro-ph.CO]}
  \BibitemShut {NoStop}%
\bibitem [{\citenamefont {Feng}\ \emph {et~al.}(2003)\citenamefont {Feng},
  \citenamefont {Rajaraman},\ and\ \citenamefont {Takayama}}]{Feng:2003xh}%
  \BibitemOpen
  \bibfield  {author} {\bibinfo {author} {\bibfnamefont {J.~L.}\ \bibnamefont
  {Feng}}, \bibinfo {author} {\bibfnamefont {A.}~\bibnamefont {Rajaraman}}, \
  and\ \bibinfo {author} {\bibfnamefont {F.}~\bibnamefont {Takayama}},\ }\href
  {\doibase 10.1103/PhysRevLett.91.011302} {\bibfield  {journal} {\bibinfo
  {journal} {Phys. Rev. Lett.}\ }\textbf {\bibinfo {volume} {91}},\ \bibinfo
  {pages} {011302} (\bibinfo {year} {2003})},\ \Eprint
  {http://arxiv.org/abs/hep-ph/0302215} {arXiv:hep-ph/0302215 [hep-ph]}
  \BibitemShut {NoStop}%
\bibitem [{\citenamefont {Steffen}(2006)}]{Steffen:2006hw}%
  \BibitemOpen
  \bibfield  {author} {\bibinfo {author} {\bibfnamefont {F.~D.}\ \bibnamefont
  {Steffen}},\ }\href {\doibase 10.1088/1475-7516/2006/09/001} {\bibfield
  {journal} {\bibinfo  {journal} {JCAP}\ }\textbf {\bibinfo {volume} {0609}},\
  \bibinfo {pages} {001} (\bibinfo {year} {2006})},\ \Eprint
  {http://arxiv.org/abs/hep-ph/0605306} {arXiv:hep-ph/0605306 [hep-ph]}
  \BibitemShut {NoStop}%
\bibitem [{\citenamefont {Benakli}\ \emph {et~al.}(2017)\citenamefont
  {Benakli}, \citenamefont {Chen}, \citenamefont {Dudas},\ and\ \citenamefont
  {Mambrini}}]{Benakli:2017whb}%
  \BibitemOpen
  \bibfield  {author} {\bibinfo {author} {\bibfnamefont {K.}~\bibnamefont
  {Benakli}}, \bibinfo {author} {\bibfnamefont {Y.}~\bibnamefont {Chen}},
  \bibinfo {author} {\bibfnamefont {E.}~\bibnamefont {Dudas}}, \ and\ \bibinfo
  {author} {\bibfnamefont {Y.}~\bibnamefont {Mambrini}},\ }\href {\doibase
  10.1103/PhysRevD.95.095002} {\bibfield  {journal} {\bibinfo  {journal} {Phys.
  Rev.}\ }\textbf {\bibinfo {volume} {D95}},\ \bibinfo {pages} {095002}
  (\bibinfo {year} {2017})},\ \Eprint {http://arxiv.org/abs/1701.06574}
  {arXiv:1701.06574 [hep-ph]} \BibitemShut {NoStop}%
\bibitem [{\citenamefont {Schissel}(2006)}]{Schissel:2006kx}%
  \BibitemOpen
  \bibfield  {author} {\bibinfo {author} {\bibfnamefont {J.}~\bibnamefont
  {Schissel}},\ }\href@noop {} {\  (\bibinfo {year} {2006})},\ \Eprint
  {http://arxiv.org/abs/hep-ph/0608014} {arXiv:hep-ph/0608014 [hep-ph]}
  \BibitemShut {NoStop}%
\bibitem [{\citenamefont {Hochberg}\ \emph {et~al.}(2014)\citenamefont
  {Hochberg}, \citenamefont {Kuflik}, \citenamefont {Volansky},\ and\
  \citenamefont {Wacker}}]{Hochberg:2014dra}%
  \BibitemOpen
  \bibfield  {author} {\bibinfo {author} {\bibfnamefont {Y.}~\bibnamefont
  {Hochberg}}, \bibinfo {author} {\bibfnamefont {E.}~\bibnamefont {Kuflik}},
  \bibinfo {author} {\bibfnamefont {T.}~\bibnamefont {Volansky}}, \ and\
  \bibinfo {author} {\bibfnamefont {J.~G.}\ \bibnamefont {Wacker}},\ }\href
  {\doibase 10.1103/PhysRevLett.113.171301} {\bibfield  {journal} {\bibinfo
  {journal} {Phys. Rev. Lett.}\ }\textbf {\bibinfo {volume} {113}},\ \bibinfo
  {pages} {171301} (\bibinfo {year} {2014})},\ \Eprint
  {http://arxiv.org/abs/1402.5143} {arXiv:1402.5143 [hep-ph]} \BibitemShut
  {NoStop}%
\bibitem [{\citenamefont {Bruggisser}\ \emph {et~al.}(2017)\citenamefont
  {Bruggisser}, \citenamefont {Riva},\ and\ \citenamefont
  {Urbano}}]{Bruggisser:2016ixa}%
  \BibitemOpen
  \bibfield  {author} {\bibinfo {author} {\bibfnamefont {S.}~\bibnamefont
  {Bruggisser}}, \bibinfo {author} {\bibfnamefont {F.}~\bibnamefont {Riva}}, \
  and\ \bibinfo {author} {\bibfnamefont {A.}~\bibnamefont {Urbano}},\ }\href
  {\doibase 10.21468/SciPostPhys.3.3.017} {\bibfield  {journal} {\bibinfo
  {journal} {SciPost Phys.}\ }\textbf {\bibinfo {volume} {3}},\ \bibinfo
  {pages} {017} (\bibinfo {year} {2017})},\ \Eprint
  {http://arxiv.org/abs/1607.02474} {arXiv:1607.02474 [hep-ph]} \BibitemShut
  {NoStop}%
\bibitem [{\citenamefont {Collar}\ and\ \citenamefont
  {Avignone}(1992)}]{Collar:1992qc}%
  \BibitemOpen
  \bibfield  {author} {\bibinfo {author} {\bibfnamefont {J.~I.}\ \bibnamefont
  {Collar}}\ and\ \bibinfo {author} {\bibfnamefont {F.~T.}\ \bibnamefont
  {Avignone}},\ }\href {\doibase 10.1016/0370-2693(92)90873-3} {\bibfield
  {journal} {\bibinfo  {journal} {Phys. Lett.}\ }\textbf {\bibinfo {volume}
  {B275}},\ \bibinfo {pages} {181} (\bibinfo {year} {1992})}\BibitemShut
  {NoStop}%
\bibitem [{\citenamefont {Collar}\ and\ \citenamefont
  {Avignone}(1993)}]{Collar:1993ss}%
  \BibitemOpen
  \bibfield  {author} {\bibinfo {author} {\bibfnamefont {J.~I.}\ \bibnamefont
  {Collar}}\ and\ \bibinfo {author} {\bibfnamefont {F.~T.}\ \bibnamefont
  {Avignone}, \bibfnamefont {III}},\ }\href {\doibase 10.1103/PhysRevD.47.5238}
  {\bibfield  {journal} {\bibinfo  {journal} {Phys. Rev.}\ }\textbf {\bibinfo
  {volume} {D47}},\ \bibinfo {pages} {5238} (\bibinfo {year}
  {1993})}\BibitemShut {NoStop}%
\bibitem [{\citenamefont {Hasenbalg}\ \emph {et~al.}(1997)\citenamefont
  {Hasenbalg}, \citenamefont {Abriola}, \citenamefont {Avignone}, \citenamefont
  {Collar}, \citenamefont {Di~Gregorio}, \citenamefont {Gattone}, \citenamefont
  {Huck}, \citenamefont {Tomasi},\ and\ \citenamefont
  {Urteaga}}]{Hasenbalg:1997hs}%
  \BibitemOpen
  \bibfield  {author} {\bibinfo {author} {\bibfnamefont {F.}~\bibnamefont
  {Hasenbalg}}, \bibinfo {author} {\bibfnamefont {D.}~\bibnamefont {Abriola}},
  \bibinfo {author} {\bibfnamefont {F.~T.}\ \bibnamefont {Avignone}}, \bibinfo
  {author} {\bibfnamefont {J.~I.}\ \bibnamefont {Collar}}, \bibinfo {author}
  {\bibfnamefont {D.~E.}\ \bibnamefont {Di~Gregorio}}, \bibinfo {author}
  {\bibfnamefont {A.~O.}\ \bibnamefont {Gattone}}, \bibinfo {author}
  {\bibfnamefont {H.}~\bibnamefont {Huck}}, \bibinfo {author} {\bibfnamefont
  {D.}~\bibnamefont {Tomasi}}, \ and\ \bibinfo {author} {\bibfnamefont
  {I.}~\bibnamefont {Urteaga}},\ }\href {\doibase 10.1103/PhysRevD.55.7350}
  {\bibfield  {journal} {\bibinfo  {journal} {Phys. Rev.}\ }\textbf {\bibinfo
  {volume} {D55}},\ \bibinfo {pages} {7350} (\bibinfo {year} {1997})},\ \Eprint
  {http://arxiv.org/abs/astro-ph/9702165} {arXiv:astro-ph/9702165 [astro-ph]}
  \BibitemShut {NoStop}%
\bibitem [{\citenamefont {Foot}(2012)}]{Foot:2011fh}%
  \BibitemOpen
  \bibfield  {author} {\bibinfo {author} {\bibfnamefont {R.}~\bibnamefont
  {Foot}},\ }\href {\doibase 10.1088/1475-7516/2012/04/014} {\bibfield
  {journal} {\bibinfo  {journal} {JCAP}\ }\textbf {\bibinfo {volume} {1204}},\
  \bibinfo {pages} {014} (\bibinfo {year} {2012})},\ \Eprint
  {http://arxiv.org/abs/1110.2908} {arXiv:1110.2908 [hep-ph]} \BibitemShut
  {NoStop}%
\bibitem [{\citenamefont {Kouvaris}\ and\ \citenamefont
  {Shoemaker}(2014)}]{Kouvaris:2014lpa}%
  \BibitemOpen
  \bibfield  {author} {\bibinfo {author} {\bibfnamefont {C.}~\bibnamefont
  {Kouvaris}}\ and\ \bibinfo {author} {\bibfnamefont {I.~M.}\ \bibnamefont
  {Shoemaker}},\ }\href {\doibase 10.1103/PhysRevD.90.095011} {\bibfield
  {journal} {\bibinfo  {journal} {Phys. Rev.}\ }\textbf {\bibinfo {volume}
  {D90}},\ \bibinfo {pages} {095011} (\bibinfo {year} {2014})},\ \Eprint
  {http://arxiv.org/abs/1405.1729} {arXiv:1405.1729 [hep-ph]} \BibitemShut
  {NoStop}%
\bibitem [{\citenamefont {Kouvaris}(2016)}]{Kouvaris:2015laa}%
  \BibitemOpen
  \bibfield  {author} {\bibinfo {author} {\bibfnamefont {C.}~\bibnamefont
  {Kouvaris}},\ }\href {\doibase 10.1103/PhysRevD.93.035023} {\bibfield
  {journal} {\bibinfo  {journal} {Phys. Rev.}\ }\textbf {\bibinfo {volume}
  {D93}},\ \bibinfo {pages} {035023} (\bibinfo {year} {2016})},\ \Eprint
  {http://arxiv.org/abs/1509.08720} {arXiv:1509.08720 [hep-ph]} \BibitemShut
  {NoStop}%
\bibitem [{\citenamefont {Foot}\ and\ \citenamefont
  {Vagnozzi}(2015)}]{Foot:2014osa}%
  \BibitemOpen
  \bibfield  {author} {\bibinfo {author} {\bibfnamefont {R.}~\bibnamefont
  {Foot}}\ and\ \bibinfo {author} {\bibfnamefont {S.}~\bibnamefont
  {Vagnozzi}},\ }\href {\doibase 10.1016/j.physletb.2015.06.063} {\bibfield
  {journal} {\bibinfo  {journal} {Phys. Lett.}\ }\textbf {\bibinfo {volume}
  {B748}},\ \bibinfo {pages} {61} (\bibinfo {year} {2015})},\ \Eprint
  {http://arxiv.org/abs/1412.0762} {arXiv:1412.0762 [hep-ph]} \BibitemShut
  {NoStop}%
\bibitem [{\citenamefont {Clarke}\ and\ \citenamefont
  {Foot}(2016)}]{Clarke:2015gqw}%
  \BibitemOpen
  \bibfield  {author} {\bibinfo {author} {\bibfnamefont {J.~D.}\ \bibnamefont
  {Clarke}}\ and\ \bibinfo {author} {\bibfnamefont {R.}~\bibnamefont {Foot}},\
  }\href {\doibase 10.1088/1475-7516/2016/01/029} {\bibfield  {journal}
  {\bibinfo  {journal} {JCAP}\ }\textbf {\bibinfo {volume} {1601}},\ \bibinfo
  {pages} {029} (\bibinfo {year} {2016})},\ \Eprint
  {http://arxiv.org/abs/1512.06471} {arXiv:1512.06471 [astro-ph.GA]}
  \BibitemShut {NoStop}%
\bibitem [{\citenamefont {Bernabei}\ \emph {et~al.}(2015)\citenamefont
  {Bernabei} \emph {et~al.}}]{Bernabei:2015nia}%
  \BibitemOpen
  \bibfield  {author} {\bibinfo {author} {\bibfnamefont {R.}~\bibnamefont
  {Bernabei}} \emph {et~al.},\ }\href {\doibase 10.1140/epjc/s10052-015-3473-y}
  {\bibfield  {journal} {\bibinfo  {journal} {Eur. Phys. J.}\ }\textbf
  {\bibinfo {volume} {C75}},\ \bibinfo {pages} {239} (\bibinfo {year}
  {2015})},\ \Eprint {http://arxiv.org/abs/1505.05336} {arXiv:1505.05336
  [hep-ph]} \BibitemShut {NoStop}%
\bibitem [{\citenamefont {Emken}\ \emph {et~al.}(2017)\citenamefont {Emken},
  \citenamefont {Kouvaris},\ and\ \citenamefont {Shoemaker}}]{Emken:2017erx}%
  \BibitemOpen
  \bibfield  {author} {\bibinfo {author} {\bibfnamefont {T.}~\bibnamefont
  {Emken}}, \bibinfo {author} {\bibfnamefont {C.}~\bibnamefont {Kouvaris}}, \
  and\ \bibinfo {author} {\bibfnamefont {I.~M.}\ \bibnamefont {Shoemaker}},\
  }\href {\doibase 10.1103/PhysRevD.96.015018} {\bibfield  {journal} {\bibinfo
  {journal} {Phys. Rev.}\ }\textbf {\bibinfo {volume} {D96}},\ \bibinfo {pages}
  {015018} (\bibinfo {year} {2017})},\ \Eprint
  {http://arxiv.org/abs/1702.07750} {arXiv:1702.07750 [hep-ph]} \BibitemShut
  {NoStop}%
\bibitem [{\citenamefont {Davis}(2017)}]{Davis:2017noy}%
  \BibitemOpen
  \bibfield  {author} {\bibinfo {author} {\bibfnamefont {J.~H.}\ \bibnamefont
  {Davis}},\ }\href {\doibase 10.1103/PhysRevLett.119.211302} {\bibfield
  {journal} {\bibinfo  {journal} {Phys. Rev. Lett.}\ }\textbf {\bibinfo
  {volume} {119}},\ \bibinfo {pages} {211302} (\bibinfo {year} {2017})},\
  \Eprint {http://arxiv.org/abs/1708.01484} {arXiv:1708.01484 [hep-ph]}
  \BibitemShut {NoStop}%
\bibitem [{\citenamefont {Kavanagh}\ \emph {et~al.}(2017)\citenamefont
  {Kavanagh}, \citenamefont {Catena},\ and\ \citenamefont
  {Kouvaris}}]{Kavanagh:2016pyr}%
  \BibitemOpen
  \bibfield  {author} {\bibinfo {author} {\bibfnamefont {B.~J.}\ \bibnamefont
  {Kavanagh}}, \bibinfo {author} {\bibfnamefont {R.}~\bibnamefont {Catena}}, \
  and\ \bibinfo {author} {\bibfnamefont {C.}~\bibnamefont {Kouvaris}},\ }\href
  {\doibase 10.1088/1475-7516/2017/01/012} {\bibfield  {journal} {\bibinfo
  {journal} {JCAP}\ }\textbf {\bibinfo {volume} {1701}},\ \bibinfo {pages}
  {012} (\bibinfo {year} {2017})},\ \Eprint {http://arxiv.org/abs/1611.05453}
  {arXiv:1611.05453 [hep-ph]} \BibitemShut {NoStop}%
\bibitem [{\citenamefont {Mahdawi}\ and\ \citenamefont
  {Farrar}(2017{\natexlab{a}})}]{Mahdawi:2017cxz}%
  \BibitemOpen
  \bibfield  {author} {\bibinfo {author} {\bibfnamefont {M.~S.}\ \bibnamefont
  {Mahdawi}}\ and\ \bibinfo {author} {\bibfnamefont {G.~R.}\ \bibnamefont
  {Farrar}},\ }\href {\doibase 10.1088/1475-7516/2017/12/004} {\bibfield
  {journal} {\bibinfo  {journal} {JCAP}\ }\textbf {\bibinfo {volume} {1712}},\
  \bibinfo {pages} {004} (\bibinfo {year} {2017}{\natexlab{a}})},\ \Eprint
  {http://arxiv.org/abs/1709.00430} {arXiv:1709.00430 [hep-ph]} \BibitemShut
  {NoStop}%
\bibitem [{\citenamefont {Mahdawi}\ and\ \citenamefont
  {Farrar}(2017{\natexlab{b}})}]{Mahdawi:2017utm}%
  \BibitemOpen
  \bibfield  {author} {\bibinfo {author} {\bibfnamefont {M.~S.}\ \bibnamefont
  {Mahdawi}}\ and\ \bibinfo {author} {\bibfnamefont {G.~R.}\ \bibnamefont
  {Farrar}},\ }\href@noop {} {\  (\bibinfo {year} {2017}{\natexlab{b}})},\
  \Eprint {http://arxiv.org/abs/1712.01170} {arXiv:1712.01170 [hep-ph]}
  \BibitemShut {NoStop}%
\bibitem [{\citenamefont {Emken}\ and\ \citenamefont
  {Kouvaris}(2017{\natexlab{a}})}]{Emken2017a}%
  \BibitemOpen
  \bibfield  {author} {\bibinfo {author} {\bibfnamefont {T.}~\bibnamefont
  {Emken}}\ and\ \bibinfo {author} {\bibfnamefont {C.}~\bibnamefont
  {Kouvaris}},\ }\href@noop {} {\enquote {\bibinfo {title} {{DaMaSCUS v1.0},
  \href{http://ascl.net/code/v/1702}{\textnormal{[ascl:1706.003]}}\textnormal{.
  Available at }\url{https://github.com/temken/damascus}},}\ } (\bibinfo {year}
  {2017}{\natexlab{a}})\BibitemShut {NoStop}%
\bibitem [{\citenamefont {Emken}\ and\ \citenamefont
  {Kouvaris}(2017{\natexlab{b}})}]{Emken:2017qmp}%
  \BibitemOpen
  \bibfield  {author} {\bibinfo {author} {\bibfnamefont {T.}~\bibnamefont
  {Emken}}\ and\ \bibinfo {author} {\bibfnamefont {C.}~\bibnamefont
  {Kouvaris}},\ }\href {\doibase 10.1088/1475-7516/2017/10/031} {\bibfield
  {journal} {\bibinfo  {journal} {JCAP}\ }\textbf {\bibinfo {volume} {1710}},\
  \bibinfo {pages} {031} (\bibinfo {year} {2017}{\natexlab{b}})},\ \Eprint
  {http://arxiv.org/abs/1706.02249} {arXiv:1706.02249 [hep-ph]} \BibitemShut
  {NoStop}%
\bibitem [{\citenamefont {Starkman}\ \emph {et~al.}(1990)\citenamefont
  {Starkman}, \citenamefont {Gould}, \citenamefont {Esmailzadeh},\ and\
  \citenamefont {Dimopoulos}}]{Starkman:1990nj}%
  \BibitemOpen
  \bibfield  {author} {\bibinfo {author} {\bibfnamefont {G.~D.}\ \bibnamefont
  {Starkman}}, \bibinfo {author} {\bibfnamefont {A.}~\bibnamefont {Gould}},
  \bibinfo {author} {\bibfnamefont {R.}~\bibnamefont {Esmailzadeh}}, \ and\
  \bibinfo {author} {\bibfnamefont {S.}~\bibnamefont {Dimopoulos}},\ }\href
  {\doibase 10.1103/PhysRevD.41.3594} {\bibfield  {journal} {\bibinfo
  {journal} {Phys. Rev.}\ }\textbf {\bibinfo {volume} {D41}},\ \bibinfo {pages}
  {3594} (\bibinfo {year} {1990})}\BibitemShut {NoStop}%
\bibitem [{\citenamefont {Albuquerque}\ and\ \citenamefont
  {Baudis}(2003)}]{Albuquerque:2003ei}%
  \BibitemOpen
  \bibfield  {author} {\bibinfo {author} {\bibfnamefont {I.~F.~M.}\
  \bibnamefont {Albuquerque}}\ and\ \bibinfo {author} {\bibfnamefont
  {L.}~\bibnamefont {Baudis}},\ }\href {\doibase 10.1103/PhysRevLett.90.221301}
  {\bibfield  {journal} {\bibinfo  {journal} {Phys. Rev. Lett.}\ }\textbf
  {\bibinfo {volume} {90}},\ \bibinfo {pages} {221301} (\bibinfo {year}
  {2003})},\ \bibinfo {note} {[Erratum: Phys. Rev. Lett.91,229903(2003)]},\
  \Eprint {http://arxiv.org/abs/astro-ph/0301188} {arXiv:astro-ph/0301188
  [astro-ph]} \BibitemShut {NoStop}%
\bibitem [{\citenamefont {Hui}\ and\ \citenamefont
  {Stewart}(1999)}]{Hui:1998dc}%
  \BibitemOpen
  \bibfield  {author} {\bibinfo {author} {\bibfnamefont {L.}~\bibnamefont
  {Hui}}\ and\ \bibinfo {author} {\bibfnamefont {E.~D.}\ \bibnamefont
  {Stewart}},\ }\href {\doibase 10.1103/PhysRevD.60.023518} {\bibfield
  {journal} {\bibinfo  {journal} {Phys. Rev.}\ }\textbf {\bibinfo {volume}
  {D60}},\ \bibinfo {pages} {023518} (\bibinfo {year} {1999})},\ \Eprint
  {http://arxiv.org/abs/hep-ph/9812345} {arXiv:hep-ph/9812345 [hep-ph]}
  \BibitemShut {NoStop}%
\bibitem [{\citenamefont {Allahverdi}\ and\ \citenamefont
  {Drees}(2002)}]{Allahverdi:2002nb}%
  \BibitemOpen
  \bibfield  {author} {\bibinfo {author} {\bibfnamefont {R.}~\bibnamefont
  {Allahverdi}}\ and\ \bibinfo {author} {\bibfnamefont {M.}~\bibnamefont
  {Drees}},\ }\href {\doibase 10.1103/PhysRevLett.89.091302} {\bibfield
  {journal} {\bibinfo  {journal} {Phys. Rev. Lett.}\ }\textbf {\bibinfo
  {volume} {89}},\ \bibinfo {pages} {091302} (\bibinfo {year} {2002})},\
  \Eprint {http://arxiv.org/abs/hep-ph/0203118} {arXiv:hep-ph/0203118 [hep-ph]}
  \BibitemShut {NoStop}%
\bibitem [{\citenamefont {Kannike}\ \emph {et~al.}(2017)\citenamefont
  {Kannike}, \citenamefont {Racioppi},\ and\ \citenamefont
  {Raidal}}]{Kannike:2016jfs}%
  \BibitemOpen
  \bibfield  {author} {\bibinfo {author} {\bibfnamefont {K.}~\bibnamefont
  {Kannike}}, \bibinfo {author} {\bibfnamefont {A.}~\bibnamefont {Racioppi}}, \
  and\ \bibinfo {author} {\bibfnamefont {M.}~\bibnamefont {Raidal}},\ }\href
  {\doibase 10.1016/j.nuclphysb.2017.02.019} {\bibfield  {journal} {\bibinfo
  {journal} {Nucl. Phys.}\ }\textbf {\bibinfo {volume} {B918}},\ \bibinfo
  {pages} {162} (\bibinfo {year} {2017})},\ \Eprint
  {http://arxiv.org/abs/1605.09378} {arXiv:1605.09378 [hep-ph]} \BibitemShut
  {NoStop}%
\bibitem [{\citenamefont {Mack}\ \emph {et~al.}(2007)\citenamefont {Mack},
  \citenamefont {Beacom},\ and\ \citenamefont {Bertone}}]{Mack:2007xj}%
  \BibitemOpen
  \bibfield  {author} {\bibinfo {author} {\bibfnamefont {G.~D.}\ \bibnamefont
  {Mack}}, \bibinfo {author} {\bibfnamefont {J.~F.}\ \bibnamefont {Beacom}}, \
  and\ \bibinfo {author} {\bibfnamefont {G.}~\bibnamefont {Bertone}},\ }\href
  {\doibase 10.1103/PhysRevD.76.043523} {\bibfield  {journal} {\bibinfo
  {journal} {Phys. Rev.}\ }\textbf {\bibinfo {volume} {D76}},\ \bibinfo {pages}
  {043523} (\bibinfo {year} {2007})},\ \Eprint {http://arxiv.org/abs/0705.4298}
  {arXiv:0705.4298 [astro-ph]} \BibitemShut {NoStop}%
\bibitem [{\citenamefont {Albuquerque}\ and\ \citenamefont {Perez de~los
  Heros}(2010)}]{Albuquerque:2010bt}%
  \BibitemOpen
  \bibfield  {author} {\bibinfo {author} {\bibfnamefont {I.~F.~M.}\
  \bibnamefont {Albuquerque}}\ and\ \bibinfo {author} {\bibfnamefont
  {C.}~\bibnamefont {Perez de~los Heros}},\ }\href {\doibase
  10.1103/PhysRevD.81.063510} {\bibfield  {journal} {\bibinfo  {journal} {Phys.
  Rev.}\ }\textbf {\bibinfo {volume} {D81}},\ \bibinfo {pages} {063510}
  (\bibinfo {year} {2010})},\ \Eprint {http://arxiv.org/abs/1001.1381}
  {arXiv:1001.1381 [astro-ph.HE]} \BibitemShut {NoStop}%
\bibitem [{\citenamefont {Mack}\ and\ \citenamefont
  {Manohar}(2013{\natexlab{a}})}]{Mack:2012ju}%
  \BibitemOpen
  \bibfield  {author} {\bibinfo {author} {\bibfnamefont {G.~D.}\ \bibnamefont
  {Mack}}\ and\ \bibinfo {author} {\bibfnamefont {A.}~\bibnamefont {Manohar}},\
  }\href {\doibase 10.1088/0954-3899/40/11/115202} {\bibfield  {journal}
  {\bibinfo  {journal} {J. Phys.}\ }\textbf {\bibinfo {volume} {G40}},\
  \bibinfo {pages} {115202} (\bibinfo {year} {2013}{\natexlab{a}})},\ \Eprint
  {http://arxiv.org/abs/1211.1951} {arXiv:1211.1951 [astro-ph.CO]} \BibitemShut
  {NoStop}%
\bibitem [{\citenamefont {Kavanagh}(2017)}]{verne}%
  \BibitemOpen
  \bibfield  {author} {\bibinfo {author} {\bibfnamefont {B.~J.}\ \bibnamefont
  {Kavanagh}},\ }\href@noop {} {\enquote {\bibinfo {title} {{verne v1.0}},}\
  }\bibinfo {howpublished} {\textnormal{. Available at
  }\url{https://github.com/bradkav/verne}, \textnormal{. Archived at }
  \href{http://dx.doi.org/10.5281/zenodo.1115601}{\textnormal{doi:10.5281/zenodo.1115601}}}
  (\bibinfo {year} {2017}),\ \Eprint {http://arxiv.org/abs/1802.005}
  {ascl:1802.005} \BibitemShut {NoStop}%
\bibitem [{\citenamefont {Abusaidi}\ \emph {et~al.}(2000)\citenamefont
  {Abusaidi} \emph {et~al.}}]{Abusaidi:2000wg}%
  \BibitemOpen
  \bibfield  {author} {\bibinfo {author} {\bibfnamefont {R.}~\bibnamefont
  {Abusaidi}} \emph {et~al.} (\bibinfo {collaboration} {CDMS}),\ }\href
  {\doibase 10.1103/PhysRevLett.84.5699} {\bibfield  {journal} {\bibinfo
  {journal} {Phys. Rev. Lett.}\ }\textbf {\bibinfo {volume} {84}},\ \bibinfo
  {pages} {5699} (\bibinfo {year} {2000})},\ \Eprint
  {http://arxiv.org/abs/astro-ph/0002471} {arXiv:astro-ph/0002471 [astro-ph]}
  \BibitemShut {NoStop}%
\bibitem [{\citenamefont {Abrams}\ \emph {et~al.}(2002)\citenamefont {Abrams}
  \emph {et~al.}}]{Abrams:2002nb}%
  \BibitemOpen
  \bibfield  {author} {\bibinfo {author} {\bibfnamefont {D.}~\bibnamefont
  {Abrams}} \emph {et~al.} (\bibinfo {collaboration} {CDMS}),\ }\href {\doibase
  10.1103/PhysRevD.66.122003} {\bibfield  {journal} {\bibinfo  {journal} {Phys.
  Rev.}\ }\textbf {\bibinfo {volume} {D66}},\ \bibinfo {pages} {122003}
  (\bibinfo {year} {2002})},\ \Eprint {http://arxiv.org/abs/astro-ph/0203500}
  {arXiv:astro-ph/0203500 [astro-ph]} \BibitemShut {NoStop}%
\bibitem [{\citenamefont {Jungman}\ \emph {et~al.}(1996)\citenamefont
  {Jungman}, \citenamefont {Kamionkowski},\ and\ \citenamefont
  {Griest}}]{Jungman:1995df}%
  \BibitemOpen
  \bibfield  {author} {\bibinfo {author} {\bibfnamefont {G.}~\bibnamefont
  {Jungman}}, \bibinfo {author} {\bibfnamefont {M.}~\bibnamefont
  {Kamionkowski}}, \ and\ \bibinfo {author} {\bibfnamefont {K.}~\bibnamefont
  {Griest}},\ }\href {\doibase 10.1016/0370-1573(95)00058-5} {\bibfield
  {journal} {\bibinfo  {journal} {Phys. Rept.}\ }\textbf {\bibinfo {volume}
  {267}},\ \bibinfo {pages} {195} (\bibinfo {year} {1996})},\ \Eprint
  {http://arxiv.org/abs/hep-ph/9506380} {arXiv:hep-ph/9506380 [hep-ph]}
  \BibitemShut {NoStop}%
\bibitem [{\citenamefont {Green}(2012)}]{Green:2011bv}%
  \BibitemOpen
  \bibfield  {author} {\bibinfo {author} {\bibfnamefont {A.~M.}\ \bibnamefont
  {Green}},\ }\href {\doibase 10.1142/S0217732312300042} {\bibfield  {journal}
  {\bibinfo  {journal} {Mod. Phys. Lett.}\ }\textbf {\bibinfo {volume} {A27}},\
  \bibinfo {pages} {1230004} (\bibinfo {year} {2012})},\ \Eprint
  {http://arxiv.org/abs/1112.0524} {arXiv:1112.0524 [astro-ph.CO]} \BibitemShut
  {NoStop}%
\bibitem [{\citenamefont {Read}(2014)}]{Read:2014qva}%
  \BibitemOpen
  \bibfield  {author} {\bibinfo {author} {\bibfnamefont {J.~I.}\ \bibnamefont
  {Read}},\ }\href {\doibase 10.1088/0954-3899/41/6/063101} {\bibfield
  {journal} {\bibinfo  {journal} {J. Phys.}\ }\textbf {\bibinfo {volume}
  {G41}},\ \bibinfo {pages} {063101} (\bibinfo {year} {2014})},\ \Eprint
  {http://arxiv.org/abs/1404.1938} {arXiv:1404.1938 [astro-ph.GA]} \BibitemShut
  {NoStop}%
\bibitem [{\citenamefont {Piffl}\ \emph {et~al.}(2014)\citenamefont {Piffl}
  \emph {et~al.}}]{Piffl:2013mla}%
  \BibitemOpen
  \bibfield  {author} {\bibinfo {author} {\bibfnamefont {T.}~\bibnamefont
  {Piffl}} \emph {et~al.},\ }\href {\doibase 10.1051/0004-6361/201322531}
  {\bibfield  {journal} {\bibinfo  {journal} {Astron. Astrophys.}\ }\textbf
  {\bibinfo {volume} {562}},\ \bibinfo {pages} {A91} (\bibinfo {year}
  {2014})},\ \Eprint {http://arxiv.org/abs/1309.4293} {arXiv:1309.4293
  [astro-ph.GA]} \BibitemShut {NoStop}%
\bibitem [{\citenamefont {Bozorgnia}\ \emph {et~al.}(2011)\citenamefont
  {Bozorgnia}, \citenamefont {Gelmini},\ and\ \citenamefont
  {Gondolo}}]{Bozorgnia:2011tk}%
  \BibitemOpen
  \bibfield  {author} {\bibinfo {author} {\bibfnamefont {N.}~\bibnamefont
  {Bozorgnia}}, \bibinfo {author} {\bibfnamefont {G.~B.}\ \bibnamefont
  {Gelmini}}, \ and\ \bibinfo {author} {\bibfnamefont {P.}~\bibnamefont
  {Gondolo}},\ }\href {\doibase 10.1103/PhysRevD.84.023516} {\bibfield
  {journal} {\bibinfo  {journal} {Phys. Rev.}\ }\textbf {\bibinfo {volume}
  {D84}},\ \bibinfo {pages} {023516} (\bibinfo {year} {2011})},\ \Eprint
  {http://arxiv.org/abs/1101.2876} {arXiv:1101.2876 [astro-ph.CO]} \BibitemShut
  {NoStop}%
\bibitem [{\citenamefont {Mayet}\ \emph {et~al.}(2016)\citenamefont {Mayet}
  \emph {et~al.}}]{Mayet:2016zxu}%
  \BibitemOpen
  \bibfield  {author} {\bibinfo {author} {\bibfnamefont {F.}~\bibnamefont
  {Mayet}} \emph {et~al.},\ }\href {\doibase 10.1016/j.physrep.2016.02.007}
  {\bibfield  {journal} {\bibinfo  {journal} {Phys. Rept.}\ }\textbf {\bibinfo
  {volume} {627}},\ \bibinfo {pages} {1} (\bibinfo {year} {2016})},\ \Eprint
  {http://arxiv.org/abs/1602.03781} {arXiv:1602.03781 [astro-ph.CO]}
  \BibitemShut {NoStop}%
\bibitem [{\citenamefont {Lewin}\ and\ \citenamefont
  {Smith}(1996)}]{Lewin:1995rx}%
  \BibitemOpen
  \bibfield  {author} {\bibinfo {author} {\bibfnamefont {J.~D.}\ \bibnamefont
  {Lewin}}\ and\ \bibinfo {author} {\bibfnamefont {P.~F.}\ \bibnamefont
  {Smith}},\ }\href {\doibase 10.1016/S0927-6505(96)00047-3} {\bibfield
  {journal} {\bibinfo  {journal} {Astropart. Phys.}\ }\textbf {\bibinfo
  {volume} {6}},\ \bibinfo {pages} {87} (\bibinfo {year} {1996})}\BibitemShut
  {NoStop}%
\bibitem [{\citenamefont {Cerdeno}\ and\ \citenamefont
  {Green}(2010)}]{Cerdeno:2010jj}%
  \BibitemOpen
  \bibfield  {author} {\bibinfo {author} {\bibfnamefont {D.~G.}\ \bibnamefont
  {Cerdeno}}\ and\ \bibinfo {author} {\bibfnamefont {A.~M.}\ \bibnamefont
  {Green}},\ }\href@noop {} {\  (\bibinfo {year} {2010})},\ \Eprint
  {http://arxiv.org/abs/1002.1912} {arXiv:1002.1912 [astro-ph.CO]} \BibitemShut
  {NoStop}%
\bibitem [{\citenamefont {Helm}(1956)}]{Helm:1956zz}%
  \BibitemOpen
  \bibfield  {author} {\bibinfo {author} {\bibfnamefont {R.~H.}\ \bibnamefont
  {Helm}},\ }\href {\doibase 10.1103/PhysRev.104.1466} {\bibfield  {journal}
  {\bibinfo  {journal} {Phys. Rev.}\ }\textbf {\bibinfo {volume} {104}},\
  \bibinfo {pages} {1466} (\bibinfo {year} {1956})}\BibitemShut {NoStop}%
\bibitem [{\citenamefont {Duda}\ \emph {et~al.}(2007)\citenamefont {Duda},
  \citenamefont {Kemper},\ and\ \citenamefont {Gondolo}}]{Duda:2006uk}%
  \BibitemOpen
  \bibfield  {author} {\bibinfo {author} {\bibfnamefont {G.}~\bibnamefont
  {Duda}}, \bibinfo {author} {\bibfnamefont {A.}~\bibnamefont {Kemper}}, \ and\
  \bibinfo {author} {\bibfnamefont {P.}~\bibnamefont {Gondolo}},\ }\href
  {\doibase 10.1088/1475-7516/2007/04/012} {\bibfield  {journal} {\bibinfo
  {journal} {JCAP}\ }\textbf {\bibinfo {volume} {0704}},\ \bibinfo {pages}
  {012} (\bibinfo {year} {2007})},\ \Eprint
  {http://arxiv.org/abs/hep-ph/0608035} {arXiv:hep-ph/0608035 [hep-ph]}
  \BibitemShut {NoStop}%
\bibitem [{ISO 2533:1975()}]{ISA}%
  \BibitemOpen
  \bibfield  {author} {ISO 2533:1975,\ }\href
  {https://www.iso.org/standard/7472.html} {\emph {\bibinfo {title} {{Standard
  Atmosphere}}}},\ \bibinfo {type} {Standard}\ (\bibinfo  {institution}
  {International Organization for Standardization},\ \bibinfo {address}
  {Geneva, CH},\ \bibinfo {year} {1975})\BibitemShut {NoStop}%
\bibitem [{\citenamefont {Lundberg}\ and\ \citenamefont
  {Edsjo}(2004)}]{Lundberg:2004dn}%
  \BibitemOpen
  \bibfield  {author} {\bibinfo {author} {\bibfnamefont {J.}~\bibnamefont
  {Lundberg}}\ and\ \bibinfo {author} {\bibfnamefont {J.}~\bibnamefont
  {Edsjo}},\ }\href {\doibase 10.1103/PhysRevD.69.123505} {\bibfield  {journal}
  {\bibinfo  {journal} {Phys. Rev.}\ }\textbf {\bibinfo {volume} {D69}},\
  \bibinfo {pages} {123505} (\bibinfo {year} {2004})},\ \Eprint
  {http://arxiv.org/abs/astro-ph/0401113} {arXiv:astro-ph/0401113 [astro-ph]}
  \BibitemShut {NoStop}%
\bibitem [{\citenamefont {Mcdonough}(2003)}]{Geochemistry}%
  \BibitemOpen
  \bibfield  {author} {\bibinfo {author} {\bibfnamefont {W.}~\bibnamefont
  {Mcdonough}},\ }\href
  {http://www.sciencedirect.com/science/referenceworks/9780080437514} {\emph
  {\bibinfo {title} {Treatise on Geochemistry}}},\ Vol.~\bibinfo {volume} {2}\
  (\bibinfo  {publisher} {Elsevier},\ \bibinfo {year} {2003})\BibitemShut
  {NoStop}%
\bibitem [{Bri(1999)}]{Britannica}%
  \BibitemOpen
  \href@noop {} {\enquote {\bibinfo {title} {{The Earth: its properties,
  composition, and structure}},}\ }\bibinfo {howpublished} {Britannica CD,
  Version 99, {Encyclop{\ae}dia Britannica, Inc.} \copyright} (\bibinfo {year}
  {1994--1999})\BibitemShut {NoStop}%
\bibitem [{\citenamefont {Rich}\ \emph {et~al.}(1987)\citenamefont {Rich},
  \citenamefont {Rocchia},\ and\ \citenamefont {Spiro}}]{Rich1987}%
  \BibitemOpen
  \bibfield  {author} {\bibinfo {author} {\bibfnamefont {J.}~\bibnamefont
  {Rich}}, \bibinfo {author} {\bibfnamefont {R.}~\bibnamefont {Rocchia}}, \
  and\ \bibinfo {author} {\bibfnamefont {M.}~\bibnamefont {Spiro}},\ }\href
  {\doibase 10.1016/0370-2693(87)90788-x} {\bibfield  {journal} {\bibinfo
  {journal} {Physics Letters B}\ }\textbf {\bibinfo {volume} {194}},\ \bibinfo
  {pages} {173} (\bibinfo {year} {1987})}\BibitemShut {NoStop}%
\bibitem [{\citenamefont {Zaharijas}\ and\ \citenamefont
  {Farrar}(2005)}]{Zaharijas:2004jv}%
  \BibitemOpen
  \bibfield  {author} {\bibinfo {author} {\bibfnamefont {G.}~\bibnamefont
  {Zaharijas}}\ and\ \bibinfo {author} {\bibfnamefont {G.~R.}\ \bibnamefont
  {Farrar}},\ }\href {\doibase 10.1103/PhysRevD.72.083502} {\bibfield
  {journal} {\bibinfo  {journal} {Phys. Rev.}\ }\textbf {\bibinfo {volume}
  {D72}},\ \bibinfo {pages} {083502} (\bibinfo {year} {2005})},\ \Eprint
  {http://arxiv.org/abs/astro-ph/0406531} {arXiv:astro-ph/0406531 [astro-ph]}
  \BibitemShut {NoStop}%
\bibitem [{\citenamefont {Erickcek}\ \emph {et~al.}(2007)\citenamefont
  {Erickcek}, \citenamefont {Steinhardt}, \citenamefont {McCammon},\ and\
  \citenamefont {McGuire}}]{Erickcek:2007jv}%
  \BibitemOpen
  \bibfield  {author} {\bibinfo {author} {\bibfnamefont {A.~L.}\ \bibnamefont
  {Erickcek}}, \bibinfo {author} {\bibfnamefont {P.~J.}\ \bibnamefont
  {Steinhardt}}, \bibinfo {author} {\bibfnamefont {D.}~\bibnamefont
  {McCammon}}, \ and\ \bibinfo {author} {\bibfnamefont {P.~C.}\ \bibnamefont
  {McGuire}},\ }\href {\doibase 10.1103/PhysRevD.76.042007} {\bibfield
  {journal} {\bibinfo  {journal} {Phys. Rev.}\ }\textbf {\bibinfo {volume}
  {D76}},\ \bibinfo {pages} {042007} (\bibinfo {year} {2007})},\ \Eprint
  {http://arxiv.org/abs/0704.0794} {arXiv:0704.0794 [astro-ph]} \BibitemShut
  {NoStop}%
\bibitem [{\citenamefont {Snowden-Ifft}\ \emph {et~al.}(1990)\citenamefont
  {Snowden-Ifft}, \citenamefont {Barwick},\ and\ \citenamefont
  {Price}}]{SnowdenIfft1990}%
  \BibitemOpen
  \bibfield  {author} {\bibinfo {author} {\bibfnamefont {D.~P.}\ \bibnamefont
  {Snowden-Ifft}}, \bibinfo {author} {\bibfnamefont {S.~W.}\ \bibnamefont
  {Barwick}}, \ and\ \bibinfo {author} {\bibfnamefont {P.~B.}\ \bibnamefont
  {Price}},\ }\href {\doibase 10.1086/185866} {\bibfield  {journal} {\bibinfo
  {journal} {The Astrophysical Journal}\ }\textbf {\bibinfo {volume} {364}},\
  \bibinfo {pages} {L25} (\bibinfo {year} {1990})}\BibitemShut {NoStop}%
\bibitem [{\citenamefont {McGuire}(1994)}]{McGuire:1994pq}%
  \BibitemOpen
  \bibfield  {author} {\bibinfo {author} {\bibfnamefont {P.~C.}\ \bibnamefont
  {McGuire}},\ }\emph {\bibinfo {title} {{Low background balloon borne direct
  search for ionizing massive particles as a component of the dark galactic
  halo matter}}},\ \href {http://wwwlib.umi.com/dissertations/fullcit?p9424987}
  {Ph.D. thesis},\ \bibinfo  {school} {Arizona U.} (\bibinfo {year}
  {1994})\BibitemShut {NoStop}%
\bibitem [{\citenamefont {Shirk}\ and\ \citenamefont
  {Price}(1978)}]{Shirk1978}%
  \BibitemOpen
  \bibfield  {author} {\bibinfo {author} {\bibfnamefont {E.~K.}\ \bibnamefont
  {Shirk}}\ and\ \bibinfo {author} {\bibfnamefont {P.~B.}\ \bibnamefont
  {Price}},\ }\href {\doibase 10.1086/155955} {\bibfield  {journal} {\bibinfo
  {journal} {The Astrophysical Journal}\ }\textbf {\bibinfo {volume} {220}},\
  \bibinfo {pages} {719} (\bibinfo {year} {1978})}\BibitemShut {NoStop}%
\bibitem [{\citenamefont {Mack}\ and\ \citenamefont
  {Manohar}(2013{\natexlab{b}})}]{Mack2013}%
  \BibitemOpen
  \bibfield  {author} {\bibinfo {author} {\bibfnamefont {G.~D.}\ \bibnamefont
  {Mack}}\ and\ \bibinfo {author} {\bibfnamefont {A.}~\bibnamefont {Manohar}},\
  }\href {\doibase 10.1088/0954-3899/40/11/115202} {\bibfield  {journal}
  {\bibinfo  {journal} {Journal of Physics G: Nuclear and Particle Physics}\
  }\textbf {\bibinfo {volume} {40}},\ \bibinfo {pages} {115202} (\bibinfo
  {year} {2013}{\natexlab{b}})}\BibitemShut {NoStop}%
\bibitem [{\citenamefont {Fan}\ \emph {et~al.}(2013)\citenamefont {Fan},
  \citenamefont {Katz}, \citenamefont {Randall},\ and\ \citenamefont
  {Reece}}]{Fan:2013yva}%
  \BibitemOpen
  \bibfield  {author} {\bibinfo {author} {\bibfnamefont {J.}~\bibnamefont
  {Fan}}, \bibinfo {author} {\bibfnamefont {A.}~\bibnamefont {Katz}}, \bibinfo
  {author} {\bibfnamefont {L.}~\bibnamefont {Randall}}, \ and\ \bibinfo
  {author} {\bibfnamefont {M.}~\bibnamefont {Reece}},\ }\href {\doibase
  10.1016/j.dark.2013.07.001} {\bibfield  {journal} {\bibinfo  {journal} {Phys.
  Dark Univ.}\ }\textbf {\bibinfo {volume} {2}},\ \bibinfo {pages} {139}
  (\bibinfo {year} {2013})},\ \Eprint {http://arxiv.org/abs/1303.1521}
  {arXiv:1303.1521 [astro-ph.CO]} \BibitemShut {NoStop}%
\bibitem [{\citenamefont {Bozorgnia}\ and\ \citenamefont
  {Bertone}(2017)}]{Bozorgnia:2017brl}%
  \BibitemOpen
  \bibfield  {author} {\bibinfo {author} {\bibfnamefont {N.}~\bibnamefont
  {Bozorgnia}}\ and\ \bibinfo {author} {\bibfnamefont {G.}~\bibnamefont
  {Bertone}},\ }\href {\doibase 10.1142/S0217751X17300162} {\  (\bibinfo {year}
  {2017}),\ 10.1142/S0217751X17300162},\ \Eprint
  {http://arxiv.org/abs/1705.05853} {arXiv:1705.05853 [astro-ph.CO]}
  \BibitemShut {NoStop}%
\bibitem [{\citenamefont {Brout}(2002)}]{Brout:2002ec}%
  \BibitemOpen
  \bibfield  {author} {\bibinfo {author} {\bibfnamefont {R.}~\bibnamefont
  {Brout}},\ }\href@noop {} {\  (\bibinfo {year} {2002})},\ \Eprint
  {http://arxiv.org/abs/gr-qc/0201060} {arXiv:gr-qc/0201060 [gr-qc]}
  \BibitemShut {NoStop}%
\bibitem [{\citenamefont {Cyburt}\ \emph {et~al.}(2002)\citenamefont {Cyburt},
  \citenamefont {Fields}, \citenamefont {Pavlidou},\ and\ \citenamefont
  {Wandelt}}]{Cyburt2002}%
  \BibitemOpen
  \bibfield  {author} {\bibinfo {author} {\bibfnamefont {R.~H.}\ \bibnamefont
  {Cyburt}}, \bibinfo {author} {\bibfnamefont {B.~D.}\ \bibnamefont {Fields}},
  \bibinfo {author} {\bibfnamefont {V.}~\bibnamefont {Pavlidou}}, \ and\
  \bibinfo {author} {\bibfnamefont {B.}~\bibnamefont {Wandelt}},\ }\href
  {\doibase 10.1103/physrevd.65.123503} {\bibfield  {journal} {\bibinfo
  {journal} {Physical Review D}\ }\textbf {\bibinfo {volume} {65}} (\bibinfo
  {year} {2002}),\ 10.1103/physrevd.65.123503}\BibitemShut {NoStop}%
\end{thebibliography}%

\end{document}